\documentclass[a4report,10pt]{article}
\usepackage[english]{babel}
\usepackage[latin1]{inputenc}
\usepackage[T1]{fontenc}
\usepackage{amsmath}
\usepackage{textcomp}
\usepackage{graphicx,rotating}
 \usepackage{epsfig}
  \usepackage{xcolor}

\newcommand{\g}{g_{D^{*-}\bar{D}^{0}\pi^-}}

\newcommand{\Dp}{{D}^+}

\newcommand{\Do}{{D}^0}
\newcommand{\Dob}{\bar{{D}}^0}

\newcommand{\pip}{\pi^{+}}
\newcommand{\pim}{\pi^{-}}
\newcommand{\pio}{\pi^{0}}

\newcommand{\Dstar}{{D}^{\ast}}

\newcommand{\Dstaro}{{D}^{\ast 0}}
\newcommand{\Dstarp}{D^{\ast +}}
\newcommand{\Dstarm}{D^{\ast -}}

\newcommand{\Vcb}{\left | {\rm V}_{cb} \right |}

\newcommand{\Bm}{{B^{-}}}
\newcommand{\Bo}{{B^{0}_d}}
\newcommand{\Bob}{\bar{B}^0_d}

\newcommand{\mumu}{\ifmmode {\mu^+\mu^-} \else ${\mu^+\mu^-} $ \fi}

\newcommand{\ba}{\begin{array}}
\newcommand{\ea}{\end{array}}
\newcommand{\bc}{\begin{center}}
\newcommand{\ec}{\end{center}}
\newcommand{\beq}{\begin{eqnarray}}
\newcommand{\eeq}{\end{eqnarray}}
\newcommand{\bes}{\begin{eqnarray*}}
\newcommand{\ees}{\end{eqnarray*}}
\newcommand{\Kz}{\ifmmode {\rm K^0_s} \else ${\rm K^0_s} $ \fi}
\newcommand{\Zz}{\ifmmode {\rm Z} \else ${\rm Z } $ \fi}
\newcommand{\qqbar}{\ifmmode {\rm q\bar{q}} \else ${\rm q\bar{q}} $ \fi}
\newcommand{\ccbar}{\ifmmode {\rm c\bar{c}} \else ${\rm c\bar{c}} $ \fi}
\newcommand{\bbbar}{\ifmmode {\rm b\bar{b}} \else ${\rm b\bar{b}} $ \fi}
\newcommand{\xxbar}{\ifmmode {\rm x\bar{x}} \else ${\rm x\bar{x}} $ \fi}
\newcommand{\rphi}{\ifmmode {\rm R\phi} \else ${\rm R\phi} $ \fi}

\usepackage{etoolbox}
\makeatletter
\patchcmd{\maketitle}{\@fnsymbol}{\@alph}{}{}  
\makeatother
  \begin{document}

\title{ \raggedleft{\small{Preprint LPT-Orsay-18-51, LAL 18-011}}
\\\strut\\\date{}
\centering{Large off-shell 
 effects in the $\bar{D}^*$ contribution to $B \to \bar{D} \pi \pi$ and  $B \to \bar{D} \pi  \bar{\ell} \nu_{\ell}$ decays.}}

\author{Alain Le Yaouanc\thanks{\bf Laboratoire de Physique Théorique (UMR8627), CNRS, Univ. Paris-Sud, Université Paris-Saclay, 91405 Orsay, France},  
Jean-Pierre Leroy$^{\mathrm a}$  and 
Patrick Roudeau\thanks{\bf Laboratoire de l'Accélérateur Linéaire, Univ. Paris-Sud, CNRS/IN2P3, Université Paris-Saclay, Orsay, France}}
\vspace{0.3cm}
\maketitle

\begin{abstract}
We stress 
 that, although the $D^*$ is very narrow (one hundred of keV), the difference between the full 
 $D^*$ contribution to $B \to \bar{D} \pi \pi$ and its zero width limit is surprisingly large~: several percents. This phenomenon is a general effect which appears when considering the production of particles that are coupled to an intermediate virtual state, stable or not, and it  persists whether the width is large or not. The effects of various cuts and of the inclusion of damping factors at the strong and weak vertices are discussed.  It is shown how the zero width limit, 
needed to compare with theoretical expectations,
can be extracted. One also evaluates the virtual $D^*_V$ contribution, which comes out roughly as found experimentally, but which is however much more dependent on cuts and uncontrollable "off-shell" effects. We suggest a way to estimate the impact of the damping factors.
\end{abstract}

\section{Motivation}

Our goal is to clarify at the same time:  

1) the theoretical meaning of the measurement of $\Gamma(B \to \bar{D}^{*} \pi)$, i.e. how one relates the direct measurements of the quantity which we shall call $\Gamma_3,$ (which is obtained  from events selected, usually, by means of a cut on the $D\pi$ mass in the 
 3-body $B \to \bar{D} \pi \pi$  process) 
  to the quantity $\Gamma_2$ which characterizes the transition  with the $D^{*}$ considered as a stable particle, which would be a purely weak process;

2) the meaning and theoretical estimate of the measurement of the so-called $D^{*}_V$ "virtual" contribution to  $B \to \bar{D} \pi \pi$. This is a complementary useful process, 
 but one whose measurement is not so well defined, and whose theoretical evaluation is less clear.

  We first consider the $\Bo \to {\bar D^0} \pi^- \pi^+$ decay channel which is simpler to interpret theoretically meanwhile our
considerations are general and we study
also $B^+ \to D^- \pi^+ \pi^+$ and semileptonic $B \to \bar{D} \pi \bar{\ell} \nu_{\ell} $ decays.

\section{The full contribution of  $\bar{D}^*$ to  $\Gamma (B \to \bar{D} \pi \pi)$ vs the zero width limit {($g^2 \to 0$)}}

The aim of this section is to display the difference between the full resonance contribution of the $\bar{D}^*$ to $B \to \bar{D} \pi \pi$  and the computation of the $B \to \bar{D}^* \pi$ decay
when the  $\bar{D}^*$ is considered as a stable particle. In this section we consider a  final state, $\Dob \pim \pip$ in which the decay $\Dstarm \to \Dob \pim$ is allowed, when using the nominal mass values of the particles involved. In section 4.3 we study the $\Dp \pim \pim$ final state in which the decay $\Dstaro \to \ \Dp \pim$ is forbidden, in the same conditions.

From now on we use the following notations\footnote{$p_1,$ $p_2$  and  $p^\prime_2$  are actually functions of $s$ but we shall usually omit to write explicitly this dependence, unless their values at different energy scales must be distinguished. We shall denote by $p_{1,D^*}$ the value of $p_1$ evaluated at the nominal mass of the resonance. The bachelor meson momentum in the $B$-meson rest system, ${p^\prime}_2$, is related to $p_2$ by ${p^\prime}_2\,=\,\,p_2\,\sqrt{s}\,/\,m_B$.}:
\begin{itemize}
\item[-] $s$: the squared invariant mass of the (would-be) resonance;
\item[-] $m_1$ and $p_1$: the mass and the modulus of the 3-momentum of the light meson stemming from the decay of the resonance (in the resonance rest system). The corresponding 4-vector is denoted by $P_1$, and a similar convention holds for the other momenta involved;
\item[-] $m_2$ and $p_2$: the mass and  the modulus of the 3-momentum of the "bachelor" light meson  (in the  resonance rest system);
\item[-] $m_{12}$: the  invariant mass of the  pair of pions.
\end{itemize}

In terms of the momenta of the various particles involved ("bachelor" $\pi^+$, final $\bar{D}$ and $\pi^-$)
 the amplitude for the decay chain $\Bo\to D^{*-}\pi^+,\,D^{*-}\to \bar{D}^0 \pi^-$ reads:

\begin{equation}
\begin{split}
{\cal M}&=\g
\,g_2 \, P_2^\mu \left[g_{\mu\nu}-\frac{(P_D+P_1)_\mu (P_D+P_1)_\nu}{s}\right] P_1^\nu\frac{1}{s-m^2_{D^*} + i \sqrt{s}\,\Gamma_{D^*}(s)}
\\=&\frac{1}{4}\frac{\g\, g_2}{s-m^2_{D^*} + i {\sqrt{s}}\,\Gamma_{D^*}(s)}  \left[\vphantom{ \frac{(m_B^2-m_2^2)(m_D^2-m_1^2)}{s}}m_B^2\,+\,m_D^2\,+  m_1^2\, +  m_2^2-s\,-2 m_{12}^2\right. \\  \strut &\mbox{\hspace{5cm}}-\left. \frac{(m_B^2-m_2^2)(m_D^2-m_1^2)}{s}\right] \label{matrix1}\\
\end{split}
\end{equation}
\\where $g_2$ takes the value:
\begin{equation}
 g_2= G_F/\sqrt{2}\,V_{ud}\,V_{cb}^*\,f_\pi\,2\, m_{D^*} a_1 A_0(m_{\pi}^2)
\label{eq:g2}
\end{equation}
in the factorization scheme \cite{Neubert:1997uc}.  As for $g\equiv \g$,
it is related to the $D^{*+}\to D^0\,\pip$ partial width through: 
\begin{eqnarray}
\Gamma_{D^{*-}\rightarrow \,\bar{D}^0\pi^-}(s)&=&\frac{g^2}{24 \pi}\frac{1}{8 s^{5/2}}\left[(s-(m_D-m_1)^2)(s-(m_D+m_1)^2)\right]^{3/2} \,F^2_R(s)\nonumber \\
 &=&\frac{g^2}{24 \pi}\frac{p_1^3}{s} \,F^2_R(s)
\label{eq:width_no_p}
\end{eqnarray}
\noindent where $F_R(s)$ is a damping factor which verifies $F_R(m^2_{D^*})\,=\,1$ (see below for details concerning those factors).
There is some arbitrariness in the form of the Breit-Wigner (see \cite{Lichard:1998ht, Isgur:1988vm}).  We stick to the  standard formulation, advocated for instance in  \cite{Patrignani:2016xqp}, Eq. (48.15), according to which the width in the denominator of the Breit-Wigner is energy-dependent. 
Thus,  $\Gamma_{D^*}(s)$  is the total width of the resonance taken at the invariant mass $\sqrt{s}$. 
This choice corresponds to what is called $BW_\delta$ in \cite{weinstein:1999} which discusses those matters in detail; $- i \sqrt{s}\, \Gamma(s)$ is precisely the absorptive part of the self-energy generated by the $D \pi$ loop calculated through Feynman graphs (see Appendix 3).

A related ambiguity occurs regarding the numerator of the resonance Breit-Wigner. In this note we use the form $g_{\mu\nu}-\frac{(P_D+P_1)_\mu (P_D+P_1)_\nu}{s}$ instead of the $g_{\mu\nu}-\frac{(P_D+P_1)_\mu (P_D+P_1)_\nu}{m^2_{D^*}}$ one suggested by the isobaric model\footnote{By isobaric model we mean effective field-theoretic models including vector fields describing spin one resonances and subject to Feynman rules, see for instance the treatment of the $\Delta$ by Gourdin and Salin \cite{Gourdin}.}.
 When estimated in terms of  resonance rest frame quantities, the expression inside the  square brackets in Eq. (\ref{matrix1}) (which stems from the first form  above) reduces to $4\,p_1\,p_2\,\cos(\theta)$\footnote{$\theta$ is the angle between the 3-momenta of the two pions, in the resonance rest-frame.}  as expected (see for example  \cite{Zemach:1965zz}  and \cite{Zemach:1968zz}) and assumed by the  experimental analyses, see in particular the $D^*_V$.

Had we used the second form, an extra term would have appeared, namely $4\,g\,g_2\,\frac{m^2_{D^*}-s}{m^2_{D^*}\,s}(m_B^2-m_2^2-\, s)$$(m_D^2-m_1^2-\, s)$. This quantity does not depend on $m_{12}$ and, consequently, will show no dependence on  $\cos(\theta).$This is due to the fact that the propagator is no longer transverse when the resonance is off-shell, i.e., it has a scalar part in addition to the spin-1 component.  The extra term vanishes at the resonance mass but could give a relatively more important contribution at the upper end of the phase-space. However this would concern the S-wave and,    since we are interested here in the P-wave channel,  we keep the other form.  

It is customary, in experimental papers, to introduce damping factors in the analyses, the so-called "Blatt-Weisskopf" functions,  although their exact  meaning  is not precisely stated. These functions have been introduced
in nuclear physics and used for particles emitted at very low momenta  within a quantum mechanical potential-well description of the nucleus; therefore
it is not clear whether they can be used in high energy reactions. In the theoretical  formula for $\cal{M}$ above this amounts to introducing  two functions $F_B(s)$ and $F_R(s)$, leading to 

\begin{equation}
\begin{split}
{\cal M^\prime}&=\frac{1}{4}\frac{g\, g_2\,F_B(s)\,F_R(s)}{s-m^2_{D^*} + i {\sqrt{s}}\,\Gamma_{D^*}(s)}  \left[\vphantom{\frac{(m_B^2-m_2^2)(m_D^2-m_1^2)}{s}}m_B^2\,+\,m_D^2\,+  m_1^2\, +  m_2^2-s\,-2 m_{12}^2 \right. \\  \strut &\mbox{\hspace{5cm}}-\left. \frac{(m_B^2-m_2^2)(m_D^2-m_1^2)}{s}\right] \label{matrixp}
\end{split}
\end{equation}

It may be reminded that the expression for $\Gamma_{D^*}(s)$ contains the term $F^2_R(s) $ (see Eq. ($\ref{eq:width_no_p})$). Those factors depend on $s$ through the momenta $p_2$ (or  $p^\prime_2$) and $p_1.$    By convention the value of the damping factors is $1$ when the resonance is "on-shell" but, as we shall see, their influence is not negligible  as one integrates the (squared) amplitude over $s$ to get $\Gamma_3$. According to Blatt and Weisskopf, for the case we are interested in of a vector  resonance, $F_R$ takes the form $F_R(s)= \sqrt{(1+( r_{BW} \,p_{1,D^*})^2)/(1+ (r_{BW} \,p_1)^2)}$. The form of $F_B$ is similar except for the substitution of $p_1$ by either $p_2$  ({\sl LHCb}) or $p^\prime_2$   ({\sl CLEO and B-factories}). This dependence introduces an extra parameter 
generically denoted by "$r_{BW}$" in the following\footnote{Actually there are two parameters, since there is no a-priori reason why the two damping factors should be identical.} and consequently an extra source of uncertainty.

Going back to  expression (\ref{matrix1}), leaving aside any contribution besides the resonance  and squaring the amplitude one gets  for the resonant contribution to the 3-body decay width:

\begin{equation}
\begin{split}
\Gamma_3\,\equiv\,\Gamma_{(\Bo\rightarrow\,D^{*-}\pi^+ ;D^{*-}\rightarrow\bar{D}^0\pi^-)}
=\frac{g^2 \,g_2^2}{(2\pi)^3}\frac{1}{256 m^3_B}&\int \frac{ds\,dm^2_{12}}{(s-m^2_{D^*} )^2+s\,{\Gamma^2_{D^*}(s)}}\,F^2_R(s)\,\times\\\quad \mbox{\hspace{5mm}}F^2_B(s)\left[\vphantom{\frac{(m_B^2-m_2^2)(m_D^2-m_1^2)}{s}}m_B^2\,+\,m_D^2\,+  m_1^2\, +  m_2^2-s\right.&\left.-2 \,m_{12}^2-\frac{(m_B^2-m_2^2)(m_D^2-m_1^2)}{s}\right]^2 \label{gamma1}
\end{split}
\end{equation}

The final integration with respect to the invariant mass of the pions leads to

\begin{equation}
\begin{split}
\Gamma_3=\frac{g_2^2 \,g^2}{192 \pi^3}\int^{(m_B-m_2)^2}_{(m_D+m_1)^2}\frac{ ds}{s^{3/2}}\,\,\frac{F^2_B(s)\,p^{\prime 3}_2(s) F^2_R(s)\,p_1^3(s)}{(s-m^2_{D^*})^2+s\,{\Gamma^2_{D^*}(s)}}
\end{split}\label{gammatot}
\end{equation}

\noindent which  can   be  rewritten as 

\begin{equation}
\Gamma_3=\frac{1}{ \pi}\int^{(m_B-m_2)^2}_{(m_D+m_1)^2} ds\, \frac{ \Gamma_{\Bo\to D^{*-} \pi^+}(s)\quad s^{1/2}\quad \Gamma_{D^{*-}\to \bar{D}^0 \pi^-}(s)}{(s-m^2_{D^*})^2+s\,\Gamma^2_{D^*}(s)}\label{gammatot2}
\end{equation}

Thus, using the general formula

$$\delta(x)= \frac{1}{\pi} \lim_{\epsilon\to 0} \,\frac{\epsilon}{x^2\,+\,\epsilon^2}$$

one immediately  gets:
\begin{equation}
\lim_{\Gamma_{D^*}\to 0} \Gamma_3= \Gamma_{\Bo\to D^{*-} \pi^+}(m^2_{D^*})\,\times\,BR
\label{eq:limgamma3}
\end{equation}
with $BR\equiv BR_{D^{*-}\to \bar{D}^0 \pi^-}({m^2_{D^*}})$ the branching ratio in the channel under consideration, taken at {\nolinebreak $\sqrt{s} = m_{D^*}$} by virtue of the $\delta$ function.

The value of the 2-body decay width is:
\begin{equation}
\Gamma_2\,\equiv\,\Gamma_{\Bo\to D^{*-} \pi^+}({m_D^*}^2) = \frac{g^2_2}{8 \pi} \frac{1}{m_{D^*}^2}\,{p^\prime_2
}^3(m^2_{D^*}).
\label{eq:gamma2}
\end{equation}

\section{Numerical aspects: dependence on the $D^*$ width}

\subsection{Dependence on $g^2$ at fixed $D^*$ mass}
In this section we measure the effect of changing the value of the $g$ coupling constant by  introducing a scaling parameter $\lambda$ so that  $g^2$  is changed into $\lambda\,\times\,g^2$ or, equivalently, $\Gamma_{D^*\,\to\,D\pi}(s)$ goes to $\Gamma_{D^*\,\to\,D\pi}(s,\lambda)\,\equiv\,\lambda\,\times\,\Gamma_{D^*\,\to\,D\pi}(s)$.  We change the {\sl total} width in the denominator in the same way so  that the partial and total widths are both scaled proportionally\footnote{Strictly speaking, this procedure is not fully correct since there is no reason why the various channels contributing to the total width should scale in the same way.}, getting:
\begin{equation}
\Gamma_3(\lambda)=\frac{1}{ \pi}\int^{(m_B-m_2)^2}_{(m_D+m_1)^2} ds\, \Gamma_{\Bo\to D^{*-} \pi^+}(s) \frac{\lambda\,s^{1/2}}{(s-m^2_{D^*})^2+\lambda^2\,s\, {\Gamma^2_{D^*}(s)}} \,\Gamma_{D^{*-}\to \bar{D}^0 \pi^-}(s)\label{gammatotlambda}
\end{equation}
 and we define ${R}(\lambda)\,\equiv\,\Gamma_3(\lambda)\,/(\Gamma_2\,\times \,BR).$ 

Letting $\lambda$ vary from 0 to 1, one should get  in the $\lambda \to 0$ limit the result announced in the preceding section (zero-width limit), $\lim_{\lambda\to 0} {R}(\lambda) = 1$   while for $\lambda = 1$ one recovers the physical situation.


\begin{figure*}[!ht]\begin{center}
\begin{tabular}{cc}
\includegraphics[scale=.7]{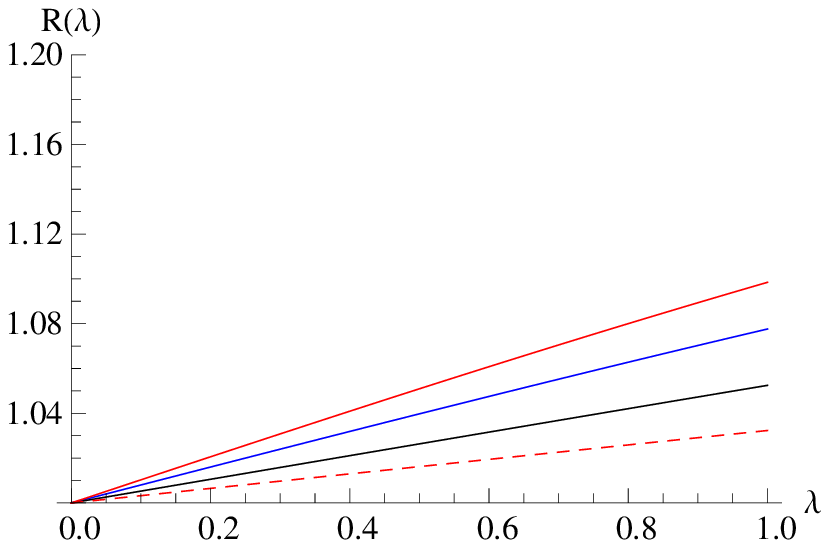}&\includegraphics[scale=.7]{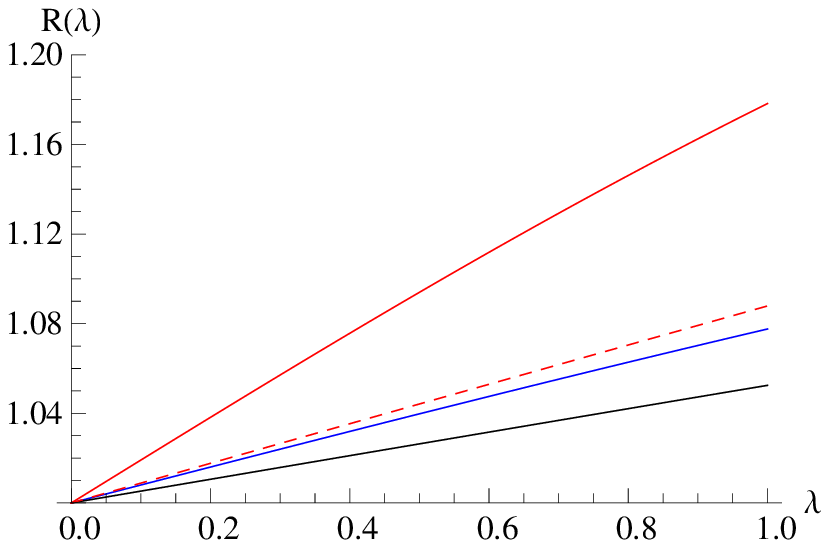}

\end{tabular}
\caption {\label{Gamma3surGamma2} {\footnotesize{\it Behavior of  ${R}(\lambda)$ as function of $\lambda$ and effect of the "Blatt-Weisskopf" damping factors:\newline\hspace*{1cm} full line, blue: without any damping factor,\newline\hspace*{1cm} full line, black: with the resonance damping only,\newline\hspace*{1cm} full line red: with the B-meson damping only, \newline\hspace*{1cm} dashed, red: with both dampings.  \newline On the left, the damping factor $F_B$ is evaluated using the momentum of the bachelor particle computed in the $B$ rest frame whereas, on the right, it is evaluated in the resonance rest frame. The parameter $r_{BW}$ is taken to be $1.6 \,GeV^{-1}$ in both cases. }}}
\end{center}\end{figure*}
\begin{figure*}[!ht]\begin{center}
\begin{tabular}{cc}
\includegraphics[scale=.7]{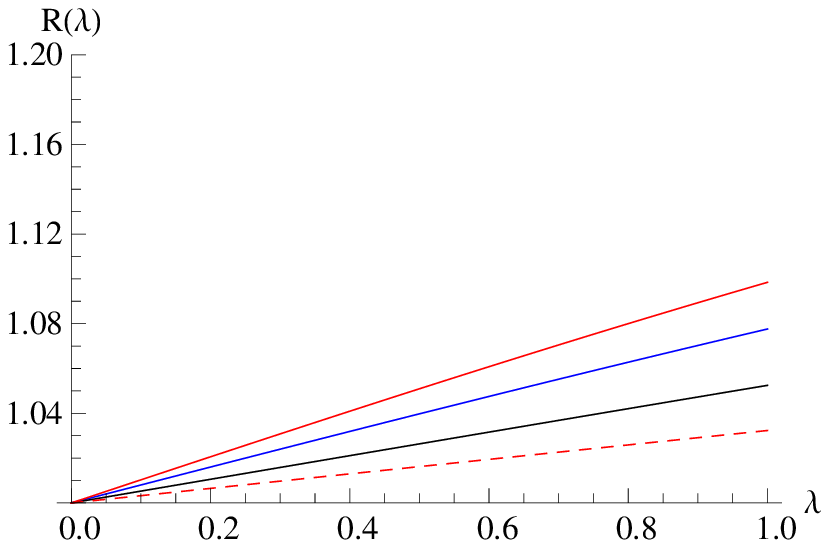}&\includegraphics[scale=.7]{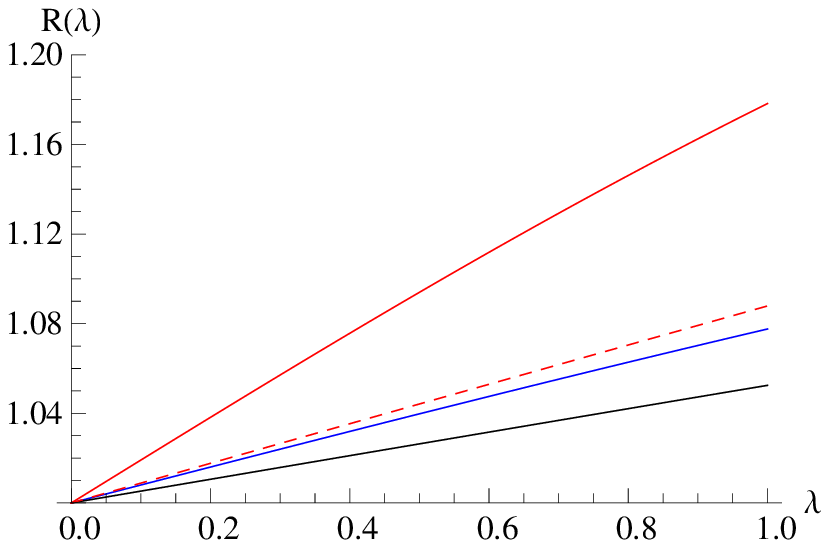}

\end{tabular}
\caption {\label{Gamma3surGamma24} {\footnotesize{\it same as Figure \ref{Gamma3surGamma2} with $r_{BW}\,=\,  4\,GeV^{-1}$ in both cases. }}}
\end{center}\end{figure*}

In Figure \ref{Gamma3surGamma2} we  show the behavior  of ${R}(\lambda)$. The numerical values are taken from the Particle Data Group Review \cite{Patrignani:2016xqp} and the coupling constants  are fitted from the two-body decay widths formulae in order to reproduce their experimental values without referring to a specific decay mechanism we get: $g=16.8$,  see Appendix 2, $a_1\,A_0\,=0.576,$ see Eq.~($\ref{eq:a1a0}$)\footnote{Note, however, that our discussion is fully independent of those numerical values.}.  It is seen that the behavior is linear and that the deviation from unity is rather large, of the order of  $10\%$ at the physical value $\lambda\,=\,1$, although the $D^*$ is still very narrow. The inclusion of the B-meson Blatt-Weisskopf factor (i.e. at the weak vertex), results in an  enhancement of the  ratio, while, on the contrary, the resonance damping induces a strong depletion. 
  We recall that the various groups (namely CLEO/B-factories and LHCb) use different  definitions  for the damping factors. Clearly, using the LHCb definition strongly  increases the effect, even though both conventions lead to   qualitatively similar effects: at the physical point ($\lambda=1$) it amounts to a several percent effect.

\subsection{Dependence on $m_{D^*}$ at fixed coupling constant}
\label{sec:fixed_coupling}
 In the previous subsection we have considered the behaviour of $ {R}(\lambda)$ at fixed $m_{D^*}$ as $g$ goes to zero. Meanwhile, 
the value of $g$ is determined by the strong interaction and is independent of the mass to first approximation. Therefore, since the nominal $D^*$ width is proportional to $g^2 p_{1,D^*}^{3}$, one has
to consider also the limit at fixed $g$, letting $p_{1,D^*}$, and consequently the width, go to zero. 
Such a limit is obtained by lowering the mass of the resonance so that
it becomes close to threshold. This corresponds to the actual situation for the $D^*$, whose narrowness is only due to the proximity of its mass  to the threshold.

\begin{figure*}[!ht]\begin{center}
\begin{tabular}{cc}
\includegraphics[scale=.7]{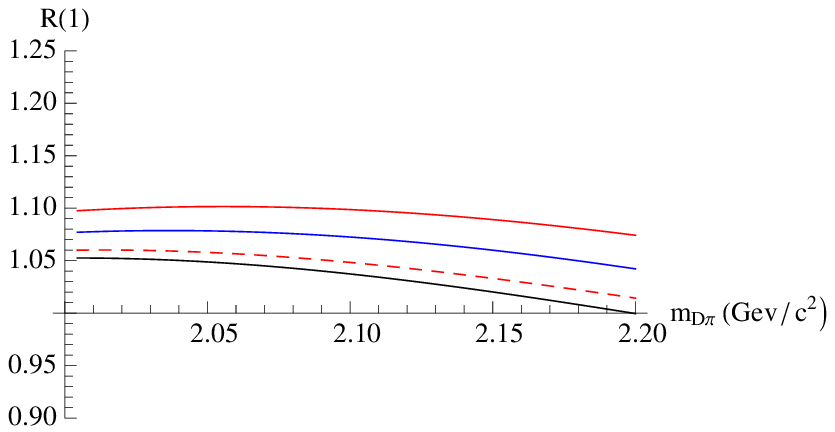}&\includegraphics[scale=.7]{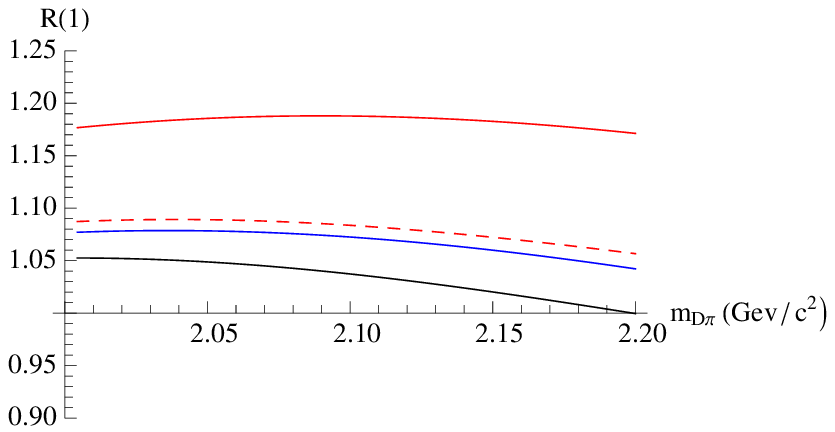}
\end{tabular}
\caption {\footnotesize{\sl{ Behavior of ${R}(1)$ as a function of the mass $m_{D\pi}$ of the resonance. The conventions are the same as in Figure $\ref{Gamma3surGamma2}.$}}}\label{Gamma3surGamma2vsm}
\end{center}\end{figure*}

\begin{figure*}[!ht]\begin{center}
\begin{tabular}{cc}
\includegraphics[scale=.7]{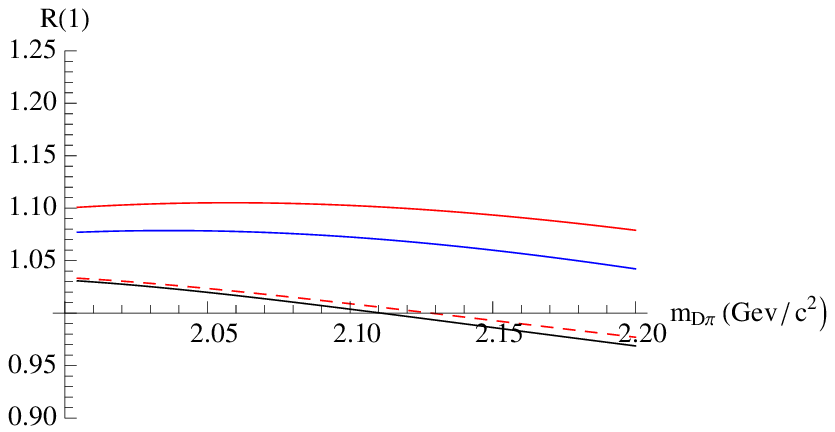}&\includegraphics[scale=.7]{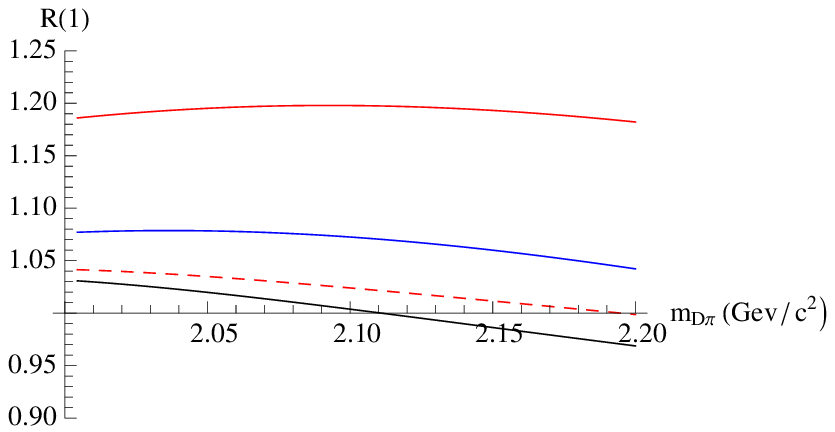}
\end{tabular}
\caption {\footnotesize{\sl{ same as  Figure $\ref{Gamma3surGamma2vsm}$ with $\,r_{BW}=\,4\,GeV^{-1}.$ }}}\label{Gamma3surGamma2vsmR4}
\end{center}\end{figure*}

Figures \ref{Gamma3surGamma2vsm} and \ref{Gamma3surGamma2vsmR4} show the behavior of  ${R}(1)$ as a function of the resonance mass. It is  seen that, whatever damping scenario  is considered, $ {R}(1 )$ remains fairly constant and significantly different from 1 when the resonance-mass varies from threshold to $2.1\, GeV,$ which corresponds to a variation of the width from $0$ to $7  MeV.$ When the mass gets close to the threshold, the low mass part of the resonance peak shrinks to $0$, which means that the departure from $0$ is mainly due to the real part of the  propagator. This is similar to the effect of the $N$-pole in $N-\pi$ scattering.

\section{Comparison with experiment in hadronic B decays}
One now turns to the question of relating above calculations
to experimental observations. As explained in previous sections, we have 
to distinguish:
 
\begin{itemize}
\item [1] the zero width 
limit $\Gamma_2$, which is a theoretical concept describing the rate 
$\Gamma_{\Bo\to \Dstarm \pip}$ as a decay to two stable particles. This is
the quantity which can be compared with corresponding theoretical
computations.
\item [2] the width obtained in 3-body decays, $\Gamma_3$, which uses $\Dob\pim\pip$ events belonging
to the decay $\Bo \to \Dstarm \pip,\,\Dstarm \to \Dob \pim$ and is obtained
by fitting the corresponding decay rate over all 
the available phase space.
\end{itemize}

In Particle Decay Tables \cite{Patrignani:2016xqp} the quantity $\Gamma_3$ is
generally used when quoting decay branching fractions of heavy mesons  into 3-body 
states, in which two of the emitted particles come from an intermediate
resonance.

The decay $\Bo\to \Dstarm \pip$ is peculiar because  a large fraction
of the $\Dob\pim$ mass distribution is concentrated over a small interval, which 
contains the $\Dstarm$ mass and, usually, only events which belong to such an interval
are selected to measure $BR(\Bo\to \Dstarm \pip)$. Unfortunately, different
experiments are using different mass intervals 
($\pm 3,\,<10 \,MeV/c^2$ by reference to $m_{D^*}$
or $m(D\pi)<2.1\,GeV/c^2$) and it is not clear  to understand, 
from present publications,
how (or even if) corrections are done, using simulated events,
to account for the presence of $D^*$ decays outside the selected range
(apart for resolution effects that are corrected). 
Therefore one needs  a precise definition of what is called a $D^*$ in
experimental measurements
to be able to combine results obtained 
in different analyses
and have a clear link with phenomenology when using simulated events. 
We detail this recommendation in Section \ref{sec:bneutral_dstar}.
It must be reminded that the $B \to \bar{D}^* \pi$ decay channel 
is used at LHC to normalize
different measurements and it is important to minimize uncertainties
on this quantity.

On the other hand the tail of the $D^*$ extends up to large $D\pi$ mass values, with distances from the pole mass that are thousands times 
larger than the width of the resonance. In effect,
as we have explained in Section 3, the behaviour of the $D^*$ tail is similar
to the one expected for other resonances, with a higher mass. It is simply the 
$D^*$ intrinsic width which is very small due to the proximity of $m_{D^*}$ 
with the decay channel threshold. Once the $D^*$ peak is eliminated by a cut
 on the $D\pi$ mass or when the $D\pi$ threshold has a higher value than
$m_{D^*}$, only the tail of the $D^*$, named $D^*_V$,  contributes in 
$D\pi\pi$ analyses. This component is usually fitted without using any
information relating its rate and mass dependence to expectations
from the $D^*$ tail. This point is discussed in Section 
\ref{sec:bneutral_dstarv}
by comparing present $D^*_V$ measurements and expectations.


\subsection{The $\Bo \to D^{*-}\pi^+$ decay channel}
\label{sec:bneutral_dstar} 

Measurements from Belle \cite{ref:btod0-belle} and BaBar 
\cite{ref:btodstar_babar_1,ref:btodstar_babar_2} collaborations 
are based on a small
fraction of their registered statistics and their results are not in good 
agreement. 

From the publications it is not clear if quoted branching fractions
are restricted to a given mass range centered on the $D^{*-}$ mass or if
measurements are corrected, using a simulation, to correspond to
$BR(\Bo \to D^{*-}\pi^+)$ over the total available phase-space?

Leaving aside these remarks and using values from \cite{Patrignani:2016xqp} we obtain:
\begin{equation}
BR(\Bo\to \Dstarm\pip)\times BR
= (1.855 \pm 0.089)\times 10^{-3}.\label{BRxBR}
\end{equation}
with $BR\,\equiv\, BR(\Dstarm \to \Dob \pim$) as in previous sections.
The value for $a_1\,A_0$, is obtained in the zero width 
approximation limit, by comparing this value to the corresponding expectation:
\begin{equation}
BR_2(\Bo\to \Dstarm\pip)\times BR= \frac{\Gamma_2\,\tau_{\Bo}}{\hbar}\,BR. 
\end{equation}
 Using the expression for $\Gamma_2$ given in Eq. (\ref{eq:gamma2}), 
this gives: 
\begin{equation}
a_1 A_0 = 0.576 \pm 0.014. 
\label{eq:a1a0}
\end{equation}
\subsubsection{Comparing our expectations and experimental results}

Taking into account the finite width of the $\Dstarm$, expected values for  
$BR(\Bo\to \Dstarm\pip)\times BR$ are obtained 
by integrating the $\Bo \to \Dob \pim \pip$ partial decay width, 
given in Eq. (\ref{gamma1}) over several $\Dob \pim$ mass intervals.
Therefore we define:
\begin{equation}
BR_3 = \frac{\Gamma_3 \, \tau_{\Bo}}{\hbar}=BR_3(m<m_{cut})+BR_3(m>m_{cut})
\end{equation}
In these evaluations,
the value of $a_1\,A_0$, obtained in the zero $\Dstarm$ width approximation,
and given in Eq. (\ref{eq:a1a0}), is used.


A relativistic Breit-Wigner distribution is used to describe the $\Dstar$
resonance:
\begin{equation}
R_{D^*}=\frac{1}{s-m^2_{D^*}+i \sqrt{s}\,\Gamma_{D^*}(s)}
\end{equation}
with
\begin{equation}
\Gamma_{D^*}(s)=\sum_{i=1}^{\infty}\Gamma_{D^*}^i \left ( \frac{p^i_1}{p^i_{1,D^*}}\right )^3 \left ( \frac{ m_{D^*}}{\sqrt{s}} \right )^2
F_R^2(p^i_1).
\label{eq:tot_width}
\end{equation}
as seen from Eqs. (\ref{matrix1}) and (\ref{eq:width_no_p}) or Eq. (47.18) of \cite{Patrignani:2016xqp}.
The value of  $m_{D^*}$ is the resonance mass and $\Gamma_{D^*}^i$ is its 
partial decay width for the $i$ channel. $p^i_1$ and $p^i_{1,D^*}$ are
the breakup momenta at the mass $m=\sqrt{s}$ and $m_{D^*}$ respectively. 
The damping factor $F_R$ is equal to unity at $m=m_{D^*}$. It decreases the
tail at large mass values of the resonance and gives some enhancement below
$m_{D^*}$. In the present analysis two parameterizations are used for the damping
factor. The one derived from a model proposed for nuclear physics by Blatt and 
Weisskopf and
another parameterization \cite{ref:bmtodp-belle}, used at B-factories in analyses
containing a $D^*$, and 
which corresponds to an exponential distribution:
\begin{equation}
F_R(p^i_1)= e^{-\alpha \left ( p^i_1-p^i_{1,D^*} \right )}.
\end{equation}
For $\Dstarp$ decays, we consider that the index $i$ varies between 1 and 3 
and corresponds 
to the channels $\Do \pip$, $\Dp \pio$, and $\Dp \gamma$ respectively. 
We have not considered additional
decay channels that should be present at high masses.
 
Results are given in Table \ref{tab:dstar}; the considered 
$m_{D\pi}$ intervals are those used in Belle \cite{ref:btod0-belle}, BaBar \cite{ref:btod0-babar} and LHCb  \cite{ref:btod0-lhcb}
in their analyses of the $\Bo\to \Dob \pim \pip$ 3-body decay channel.  Values considered for $r_{BW}$ or $\alpha$ are  representative of those measured in different experiments, as indicated in the last column of Table \ref{tab:b0todpipi}.


\begin{table}[!htb]
\begin{center}
  \begin{tabular}{|c|c|c|c|}
    \hline
$r_{BW}\,{\rm or}\,\alpha\,(GeV/c)^{-1}$   & $0$ & $1.6$& $4.0$\\
   \hline
no mass cut & $1.998$ & $1.930$& $1.890$\\
            & $1.998$ & $1.966$& $1.914$\\
   \hline
$\Delta m <3\,MeV/c^2$& $1.840$ & $1.840$& $1.840$\\
                       & $1.840$ & $1.840$& $1.840$\\
   \hline
$\Delta m<10\,MeV/c^2$& $1.855$ & $1.854$& $1.853$\\
                      & $1.855$ & $1.855$& $1.855$\\
   \hline
$m(\Dob \pi^-)<2.1\,GeV/c^2$& $1.887$ & $1.880$& $1.873$\\
                           & $1.887$ & $1.886$& $1.882$\\
\hline
  \end{tabular}
  \caption[]{\it { Values for $BR_3(m<m_{cut})=BR(\Bo\to D^{*-}\pi^+)\times BR\times 10^3$ obtained for different choices of the  mass range 
around the $\Dstarm$ mass and using an exponential (first line) or
the Blatt-Weisskopf parameterization (second line) for the damping factors. 
The theoretical expression, obtained in the zero width approximation, is
normalized to data to fix the value of the parameter $a_1 \,A_0$.
The value for $BR_2(\Bo\to \Dstarm\pip)\times BR\, is\, 
{\it 1.855 \times 10^{-3}}$ (Eq.(\ref{BRxBR})).
It can be noted that  $BR_3(m<m_{cut})$ branching fractions are almost 
independent of the value 
of the damping parameter, $r_{BW}$ or $\alpha$, once the measurement is done within a given mass range, meanwhile their values depend on the chosen mass interval.}
  \label{tab:dstar}}
\end{center}
\end{table}

When integrating over the whole Dalitz plane (second line), 
the expected branching fraction
decreases by about $5\%$ when $\alpha$ varies between $0$ and $4\,(GeV/c)^{-1}$.
This variation is reduced below the 2 permil level if, for example, a mass range
of $\pm 10 \,MeV/c^2$ is used to select $\Dstarm$ candidates.

Therefore, if the $\Dstarm$ production is measured within a fixed 
mass range, around the $\Dstarm$ mass, comparison with theoretical
expectations, obtained in the same conditions, can be of high accuracy
and are not dependent on the parameterization of damping form factors.

Ratios between expected widths in different mass
intervals and the value obtained in the narrow width approximation
are independent of $a_1 \, A_0$. 
\begin{equation}
{R}(m_{cut})= \frac{\Gamma_3(m<m_{cut})}{\Gamma_2 \times BR}
\end{equation}

Without any cut on the $\Dob \pim$ mass, this 
ratio changes from $1.077$ if no damping form factors are included
and $1.020$ using form factors with an exponential dependence and 
$\alpha=4\,(GeV/c)^{-1}$. This variation comes from the tails in the mass
distribution, outside the $\Dstarm$ region. 
Restricting the mass interval to $\Delta m<10\,MeV/c^2$,
the ratio is equal to unity and variations observed by considering
different hypotheses on damping factors are at the permil level.

We note also a variation of $2.5\%$ on the value of the branching fraction
when considering the three mass intervals  given in Table 
\ref{tab:dstar} and used by different experiments. 
This quantifies the importance of quoting the limits of the 
 $\Delta m$ interval over which the branching fraction is evaluated
by the various analyses.

It is also possible to define the cut ($m_{cut}^0$) on the $\Dob \pim$ mass so that
the corresponding integrated three body decay branching fraction corresponds to
the value expected from theory in the zero width approximation. It is 
independent of the value of the form factor $a_1\,A_0$ and almost also
of the damping factors:

\begin{equation}
m_{cut}^0=m_{D^*}+(9-10)\, MeV/c^2
\label{eq:mcut}
\end{equation}

These results are obtained with the momentum of the bachelor pion,
which enters in $F_B(p)$, computed in the $B$ meson rest frame, as was done
at B-factories. This aspect is developed in section \ref{sec:bneutral_dstarv}.

\subsubsection{Proposal to quote $BR(\Bo \to \Dstarm \pip)$}

To avoid uncertainties 
related to the  unknown shape of damping form factors
and to account for effects related to the choice of the $m_{cut}$ value, 
we advocate to quote $BR(\Bo \to \Dstarm \pip)$ for events
selected within a specified $m_{D\pi}$ interval. Measured quantities have to 
be corrected for different experimental 
effects,
 using simulated events, but no correction must be applied to account
for the cut on $m_{D\pi}$ (apart for resolution effects) so that 
corrected events correspond only to those situated in the quoted mass interval
before any experimental effect.

If experiments use different intervals in $m_{D\pi}$ it is necessary
to correct individual measurements so that they correspond to the same mass 
range, before computing the average.

The obtained value will then  be essentially independent of hypotheses for
damping factors if the combinatorial background, present under the 
$D^*$, in the selected mass interval, can be estimated in a way which does not
depend much on the high mass tail of the signal.  To compare with theory, the value $m^0_{cut}$, given in Eq. (\ref{eq:mcut}), is adequate.

\subsection{Rate and branching fraction for the virtual contribution $\Bo \to \bar{D}_V^{*-}\pip$}
\label{sec:bneutral_dstarv}
The measured fraction of $\Bo \to \bar{D}_V^{*}\pip$ events in the 3-body
$\Bo \to \Dob \pim \pip$ final state, after vetoing the $\Dstar$ mass region
($m_{D\pi}>m_{cut}$), is of the order of $10\,\%$ and is
concentrated at low $ \Dob \pim$ mass values.

\subsubsection{Theoretical expectations for the  $D^{*-}_V$ component}

In Table \ref{tab:dstarv}, values for  
$BR_3(m>m_{cut})=BR(\Bo \to D^{*-}_V\pip)\times BR(D_V^{*-}\to \Dob\pim)$
are obtained using the value of $a_1 \,A_0$ previously determined and for two 
parameterizations of damping form factors. In the following we use the notation~: $BR_V\,\equiv BR(D_V^{*-}\to \Dob\pim)$ because this quantity can have a value different from $BR$, which was defined at the resonance mass.




\begin{table}[!htb]
\begin{center}
  \begin{tabular}{|c|c|c|c|c|c|}
    \hline
$r_{BW}\,{\rm or}\,\alpha\,(GeV/c)^{-1}$    & $0.0$ & $1.6$& $3.0$& $4.0$& $5.0$\\
   \hline
$\Delta m >3\,MeV/c^2$& $1.578$ & $0.899$& $0.615$& $0.499$& $0.420$\\
                      &         & $1.256$& $0.893$& $0.742$& $0.638$\\
\hline
                      &         & $1.769$& $1.131$& $0.900$& $0.751$\\
\hline   \hline
$\Delta m>10\,MeV/c^2$& $1.427$ & $0.753$& $0.476$ & $0.363$& $0.288$\\
                      &        & $1.105$& $0.743$ & $0.594$& $0.492$\\
\hline
                      &        & $1.617$& $0.980$ & $0.751$& $0.604$\\
\hline\hline
$m(\Dob \pi^-)>2.1\,GeV/c^2$& $1.111$ & $0.494$& $0.254$& $0.165$& $0.110$\\
                            &        & $0.797$& $0.456$& $0.326$& $0.244$\\
\hline
                      &        & $1.297$& $0.682$ & $0.473$& $0.346$\\
\hline

  \end{tabular}
  \caption[]{\it {$BR_3(m>m_{cut})=BR(\Bo\to D_V^{*-}\pi^+)\times BR_V\times 10^4$ expectations 
for different values of the damping parameter and of the  
selected mass range. For each mass range, the first line corresponds to 
the exponential parameterization of the damping form factor,  
the second line is obtained with the Blatt-Weisskopf parameterization
and the third line uses the same parameterization but the bachelor pion momentum
is computed in the resonance rest frame.}
  \label{tab:dstarv}}
\end{center}
\end{table}

Results given in the first two lines, for each mass range, 
are obtained using the value of the bachelor pion momentum,
which enters in the damping factor $F_B(p)$,
computed in the $B$ rest frame. If, instead, we use the corresponding momentum 
value obtained in the $D\pi$ rest frame we get the results given in the third line. In this case, one notes that, for $r_{BW}=1.6\,(GeV/c)^{-1}$, branching fractions are higher than
without damping. This effect was also apparent in Figure \ref{Gamma3surGamma2}. Such
differences are obtained using the Blatt-Weisskopf parameterization and
we observe that using an exponential distribution gives much more dramatic 
differences: the $D^*_V$ component increases by more than one hundred times.
These effects are not usually mentioned in publications because they are
not present, neither in B-factories analyses, as they take the bachelor pion 
momentum evaluated in the $B$ rest frame, nor in LHCb which uses the resonance
rest-frame but does not use any exponential form factor distribution. 
It can be shown that, if the bachelor pion momentum is evaluated in the
$B$ rest frame, then the product $F_R(p_1) \times F_B(p^{\prime}_2)$ goes to one
for large $D\pi$ masses (if the same function is used for $F_R$
and $F_B$) whereas it can take arbitrary large values if $p_2$
is evaluated in the resonance rest frame.

Let us recall that there are no really compelling theoretical arguments for the introduction of the Blatt and Weisskopf damping factors, and even less for choosing such or such momentum dependence. However results are sensitive to them as can be concluded, for instance, from Table \ref{tab:dstarv} and this constitutes a source of uncertainty. Our present conclusion, considering this arbitrariness in the 
parameterization of damping factors, is to consider that the bachelor pion
momentum, that enters in $F_B$, has to be evaluated in the $B$ rest frame.
If the value of the damping parameter, $r_{BW}$, used in $F_B$, is smaller than
the one that enters in $F_R$, the total damping will be lower than unity
at large $m_{D\pi}$.
This indicates also that dedicated studies are needed to measure directly
these form factors.  

\subsubsection{Experimental measurements of the $D^{*-}_V$ component} 

Measurements obtained by Belle, BaBar and LHCb collaborations are compared
with expectations in Table \ref{tab:b0todpipi} and in Figure \ref{fig:theorie_expt}. These values are extracted from Table \ref{tab:dstarv} using corresponding values for $r_{BW}$ and $\alpha$.



\begin{table}[!htb]
\begin{center}
{\scriptsize
  \begin{tabular}{|c|c|c|c|}
    \hline
 Experiment  & $BR(\Bo \to D^{*-}_V\pip)$ & our evaluation& $r_{BW}\,{\rm or}\,\alpha$\\
  & $ \times BR(D^{*-}_V \to \Dob \pim) \times10^4$ & (exponential/Blatt-Weisskopf)&$(GeV/c)^{-1}$\\
\hline
 Belle \cite{ref:btod0-belle}& $0.88\pm 0.13\,(no \, syst.)$ & $0.90^{+0.68}_{-0.28}\,/\,1.26^{+0.32}_{-0.37}$ & $ 1.6^{+1.4}_{-1.6}$\\
\hline
BaBar \cite{ref:btod0-babar}& $1.39\pm0.08 \pm0.16 \pm0.35 \pm0.02$&  $0.36^{+0.12}_{-0.07}\,/\,0.59^{+0.15}_{-0.10}$ & $4 \pm 1$\\ 
\hline
LHCb \cite{ref:btod0-lhcb}& $0.78\pm0.05\pm0.02\pm0.15$& $0.49\,/\,0.79$ &$1.60 \pm 0.25$ \\
\hline
  \end{tabular}
}
  \caption[]{\it {Measurements of $D^*_V$ components
in $\Bo \to \Dob \pim \pip$ decays are compared
with expectations. 
The latter are provided
for two choices of the damping factor parameterization, exponential
and Blatt-Weisskopf respectively and
using central values and uncertainties on $\alpha$ or $r_{BW}$ quoted by corresponding
experiments (apart for Belle for which we use a variation between 0 and 3 $(GeV/c)^{-1}$).}
  \label{tab:b0todpipi}}
\end{center}
\end{table}

In the Belle analysis, only statistical uncertainties were quoted.
The variation range for $r_{BW}$ (and $\alpha$), between 0 and $3\,(GeV/c)^{-1}$ is chosen
to illustrate the sensitivity of theoretical expectations on the value
of this parameter. 

In the BaBar measurement, 
the dominant uncertainty comes from the parameterization of the $\bar{D}^0\pim$ S-wave, in the threshold region,
including a ``dabba'' component.
 
In the LHCb measurement, the quoted uncertainty on $r_{BW}$ is very small
when compared with previous determinations, meanwhile it does not include
any systematic uncertainty on this parameter\footnote{The value of $r_{BW}$ measured by LHCb cannot be directly compared with previous determinations because, in LHCb, the damping $F_B(p)$ is evaluated using the momentum (p) computed in the resonance rest frame instead of using the B rest frame.}.

From Table \ref{tab:b0todpipi} it appears that measured and expected values
for the $D^*_V$ component are compatible, as already
observed by Belle \cite{ref:btod0-belle}. Meanwhile 
experimental uncertainties remain quite large (those from LHCb being 
underestimated) and  are difficult to estimate
because they are mainly of theoretical origin, being dependent on the assumed
value for $r_{BW}$ (or $\alpha$) and on hypotheses for the variation of the damping factor with 
$m_{D\pi}$. It must be noted also that the value of the 
$D^*_V$ component is dominated by the low mass region.

\subsubsection{Expected variation of the $D^{*-}_V$ component with $m_{D\pi}$} 
Experiments have usually assumed a relativistic Breit-Wigner distribution
for the  $D^{*-}_V$ component (Belle, BaBar). In the
LHCb analysis \cite{ref:btod0-lhcb}, an arbitrary distribution
is fitted on data:
\begin{equation}
R(s)= e^{-\beta_1 (s-5.4)-i\beta_2(s-5.7)}.
\label{eq:rm_lhcb}
\end{equation}
This distribution has two problems to describe a $D^{*-}_V$ component: a 
very fast fall-off versus $m_{D\pi}$ and an unexpected phase variation
(the $D^{*-}_V$ amplitude is expected to be real and the phase to be constant, 
away from $m_{D^*}$).
But no experiment has really measured the $D^{*-}_V$ lineshape.

\begin{figure}[htbp]
  \begin{center}
     \epsfig{file=
     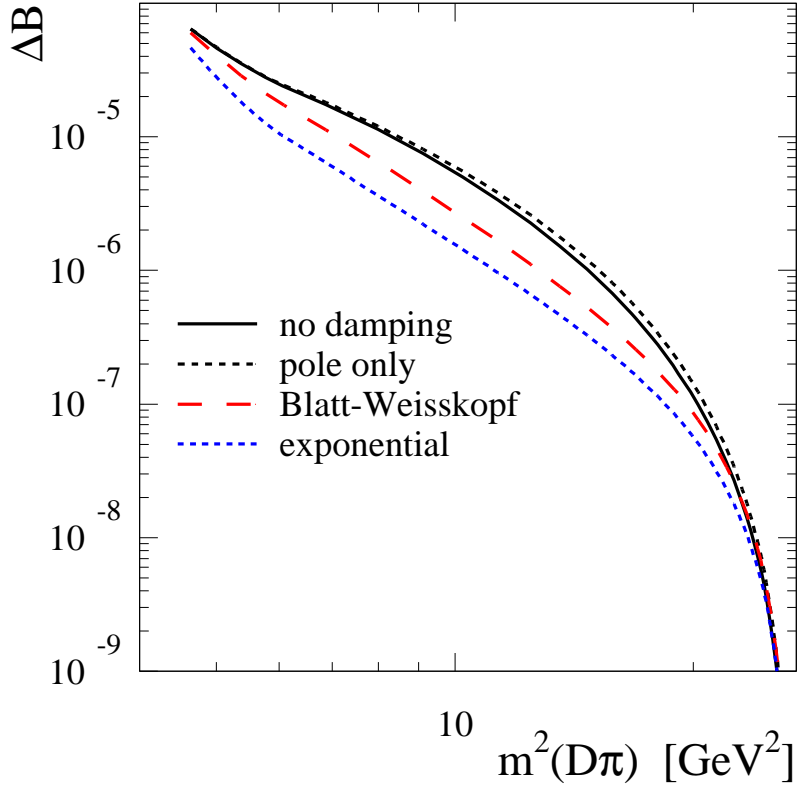,width=\textwidth} 
  \end{center}
 \caption[]{\it Comparison between the expected variation of the $D_{V}^{*-}$ component versus $m^2(D\pi)$ for different choices of damping form factors. The black full line is obtained without damping whereas Blatt-Weisskopf (red line) and exponential (blue line) damping factors are used, with the same value for the  parameter ($\alpha$ or $r_{BW}=1.6\,(GeV/c)^{-1}$). The black dashed line is obtained assuming that the total $\Dstar$ decay width is equal to zero (therefore, in this case, the $\Dstar$ amplitude is of course real). Some difference is observed at large masses ($m^2(D\pi)>10\,(GeV/c^2)^2$) which becomes non-visible, once some damping is present. $\Delta B$ is the expected branching fraction in a bin. There are 20 equal-size bins between $(2.02\, GeV/c^2)^2$ and $(m_B-m_{\pi})^2$.}   
\label{fig:width_effect}
\end{figure}

It has to  be noted that the expected mass distribution is almost independent of the exact value
of the $\Dstar$ total decay width. This is illustrated in Figure
\ref{fig:width_effect} from which it can be concluded that the 
$D^*_V$ mass distribution is the one expected from a simple pole, modified
by damping form factors.

\begin{figure}[htbp]
  \begin{center}
     \epsfig{file=
     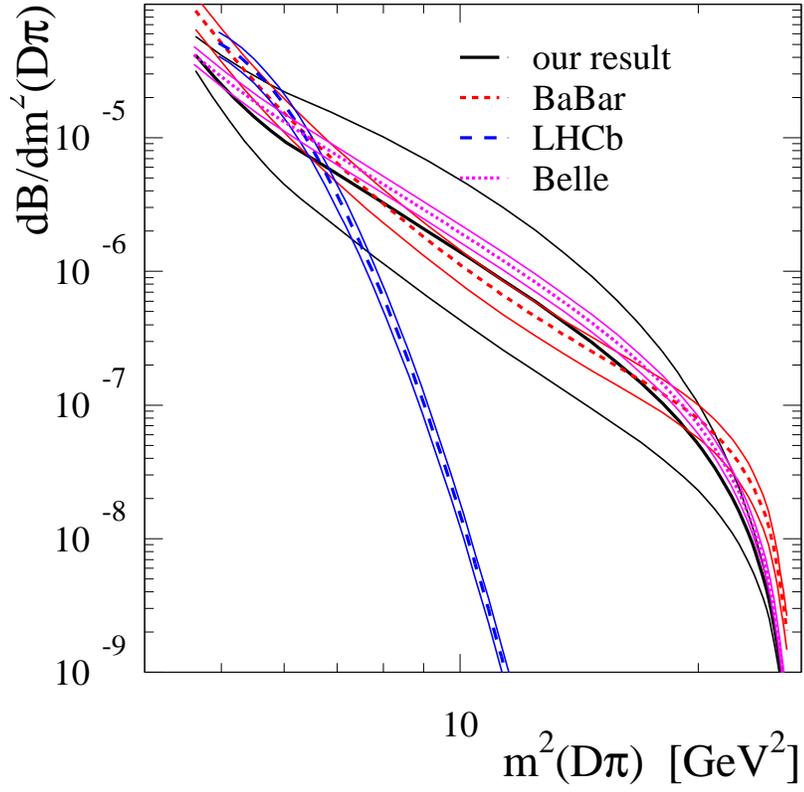,width=\textwidth}
  \end{center}
 \caption[]{\it Comparison between fitted $D_{V}^{*-}$ components, in BaBar, Belle and LHCb (dashed lines), with our expectation. For the latter,
we use exponential damping factors with $\alpha=1.6\,(GeV/c)^{-1}$ (full line), while the thin black lines on each side of the expectation are obtained by changing the value of $\alpha$ between $0.$ and $3.\,(GeV/c)^{-1}$. Thin lines drawn for each of fitted
experimental curves correspond to quoted uncertainties on $D^*_V$, given in publications.}  
\label{fig:theorie_expt}
\end{figure}

We display, in Figure \ref{fig:theorie_expt}, comparisons between $D^*_V$ distributions
fitted by experiments and our expectations. The latter are  obtained with 
the exponential parameterization of damping factors and we use $\alpha=1.6\,(GeV/c)^{-1}$.

It has to be reminded that our evaluations are based on the $D^*$
production in the mass region of the resonance and are therefore absolutely
normalized. 
The distribution obtained in Belle is compatible with our expectation.
The agreement in rate is not trivial. Meanwhile, for the mass variation,
we have used the same parameterization (exponential with $\alpha =1.6\,(GeV/c)^{-1}$)
as favored by Belle.
BaBar and LHCb observe a higher rate at low mass values. 

Because the distribution is essentially fixed by the $D^*$ pole, 
even in the presence of damping factors, we consider that the $D^*_V$
component has a non negligible contribution at large masses. Therefore
the fitted distribution by LHCb, with a fast fall-off, is not physical. 




\subsection{The $B^- \to D^+ \pim \pim$ decay channel}
\label{sec:bcharged}
The LHCb collaboration has obtained a high statistics measurement 
of the decay $B^- \to D^+ \pim \pim$ \cite{ref:bmtodp-lhcb}. 
Previous compatible results were obtained by Belle \cite{ref:bmtodp-belle}
and BaBar \cite{ref:bmtodp-babar} collaborations
but systematic uncertainties were not provided on the $D^*_V$ component.
Experimentally this channel
has the interest, when compared with $B^0 \to \Dob \pim \pip$ that,
the $\pim \pim$ final state being exotic, the decay amplitude 
is easier to parameterize  and the analysis is more sensitive to the various components in the $D\pi$
final state. Meanwhile, for theory, this decay is more difficult to interpret, 
being of Class III. But, independently
of any theoretical prejudice, it is possible to verify if the measured
$D^*_V$ component:
\begin{equation}
BR(B^- \to D^{*0}_V \pim)\times BR(D^{*0}_V  \to D^+ \pim)
=(1.09 \pm 0.07 \pm 0.07 \pm 0.24 \pm 0.07)\times 10^{-4}
\label{eq:bm_data}
\end{equation}
is compatible with the tail expected from the $D^{*0}$.

In this comparison we use the measured contribution of the $\Dstaro$  
in the decay  
$B^- \to\Dstaro \pim,\,\Dstaro \to D^0 \pio$, with a branching
fraction equal to
$(4.90\pm0.17)\times 10^{-3}\times (64.7\pm0.9)\times 10^{-2}=(3.17\pm0.12)\times 10^{-3} $ \cite{Patrignani:2016xqp}.

We have computed the corresponding decay rate by integrating 
the square of the decay amplitude modulus,
given in Eq.~(\ref{eq:ampli_bm}), 
over the $D^0 \pio \pim$ phase space, restricting the $D^0 \pio$ mass interval
to values below $m_{D\pi}<2.020\,GeV/c^2$ to isolate the $\Dstaro$ meson.

\begin{equation}
{\cal A}=C \,R_{D^*}(m) \, F_B(p_2^{\prime}) F_R(p_1) T_1(p_2,p_1,\cos{(\theta)})
\label{eq:ampli_bm}
\end{equation}

As already discussed in Section \ref{sec:bneutral_dstarv}, this expression differs from the one used by LHCb in the evaluation of
the damping $F_B$ for which we take the momentum of the bachelor pion computed
in the $B$ rest frame in place of the resonance frame.

The value of the normalization factor (noted $C$) is then determined 
such that this
evaluation corresponds to the measured value.

To obtain the $D^*_V$ contribution in the $B^- \to D^+ \pim \pim$ 
decay channel we assume that it comes from the decay chain:
$B^- \to\Dstaro \pim,\,\Dstaro \to D^+ \pim$.  
The decay threshold having a higher value than $m_{\Dstaro},$ it is not possible
to compute the value of the decay momentum, at the resonance mass, which enters
 in the expression of the partial decay width given in Eq. (\ref{eq:tot_width}).
In such circumstances, usually, an effective mass is introduced in published
analyses, which has a value much higher than the threshold. Measurements
of the corresponding $D^*_V$ component are essentially independent of this
choice, mainly because fractions and not absolute decay rates are measured.
In practice, if one takes the expression for the mass dependent decay
width, as given in Eq. (\ref{eq:width_no_p}) which does not refer to the decay
width at the resonance mass, it is not needed to use any effective mass.
As for damping factors, we take them equal to unity at the decay threshold.

The decay amplitude is symmetrized because there are two possible
$D^+\pim$ mass values, noted respectively $m_{min.}$ and  
$m_{max.}$.

\begin{equation}
A_{D^*_V}={\cal A}(m_{min.})+{\cal A}(m_{max.})
\label{eq:ampli}
\end{equation}
where the amplitude ${\cal A}(m)$ is given in Eq.~(\ref{eq:ampli_bm}).
The expected decay rate is obtained by integrating $C^2 |A_{D^*_V}|^2$ over
the plane defined by the variables $m_{min.}^2$ and  $m_{max.}^2$.

The values  given in Table \ref{tab:d-pipi} are obtained for different hypotheses on the $\alpha$ or  $r_{BW}$ 
parameters and using the exponential and the Blatt-Weisskopf parameterizations
for $F_{B,R}$.


\begin{table}[!htb]
\begin{center}
  \begin{tabular}{|c|c|c|c|c|c|}
    \hline
$r_{BW}\,{\rm or}\,\alpha\,(GeV/c)^{-1}$    & $0.0$ & $2.0$& $3.0$& $4.0$& $5.0$\\
   \hline
$BR(B^- \to D^{*0}_V \pim)$& $2.81$ & $0.68$& $0.43$& $0.30$& $0.22$\\
$\times BR(D_V^{*0}\to D^+\pi^-) \times 10^4$  & $2.81$ & $1.77$& $1.27$& $0.98$& $0.78$\\
  & $2.81$ & $2.86$& $1.89$& $1.37$& $1.06$\\
 \hline
  \end{tabular}
  \caption[]{\it { $BR(B^-\to D_V^{*0}\pi^-)\times BR(D_V^{*0}\to D^+\pi^-)\times 10^4$ expectations 
for different values of the damping parameter. 
 The bachelor pion momentum,  entering in the $F_B$ damping, is computed in the $B$ rest frame in the second and third lines, and in the resonance frame  in the last one. In the second line the exponential parameterization is used whereas for the two other lines we take the Blatt-Weisskopf expression.}
  \label{tab:d-pipi}}
\end{center}
\end{table}
{Using the parameterization 
of LHCb with $r_{BW}=(4\pm 1)\,( GeV/c)^{-1}$,  which is the value they assume for this parameter, their measurement in Eq. (\ref{eq:bm_data}) has to be compared with our estimate given in the last  line of Table \ref{tab:d-pipi}: $1.4^{+0.6}_{-0.3}\times 10^{-4}.$

\section{The $\bar{B}\to [D \pi] \ell \bar{\nu}_{\ell}$ final state}
\label{sec:bsemilept}
Similarly to what we have done  for hadronic decays, we consider two regions in the $D^* \to D\pi$ mass 
distribution.
The low mass region is used to measure the $D^*$ component which plays an important role
in the determination of the $\Vcb$ parameter. At higher masses, the tail of the
$D^*$ mass distribution is noted  $D^*_V$, as in previous sections.
The component, denoted as $[D \pi]_{broad}$, corresponds to experimental measurements of $D\pi$ final states, after a cut on $m_{D\pi}$ and 
from which $D_2^*\to D\pi$ decays are subtracted.  
The interpretation of these $[D \pi]_{broad}$ events in terms
of physical components  has been problematic for a long time. It has been  most often considered that 
they are
coming from  $D_0^*  \to D \pi$ decays but this has not been established
experimentally and does not agree with theoretical predictions
\cite{ref:bdpidl_th}. 
From theory it is expected that narrow states are produced at a larger rate than
broad states because $\tau_{3/2}(1)>\tau_{1/2}(1)$, where $\tau_{3/2}(w)$ and 
 $\tau_{1/2}(w)$ are the Isgur-Wise form factors \cite{Isgur:1991} and $w$ is the product of the 4-velocities of the $B$ and $D$ mesons, and additionally because of kinematical factors. Numerically, the expected 
branching fractions are an order of magnitude higher for narrow states
whereas the experimental value:
\begin{equation}
BR(\bar{B}^0_d \to [D \pi]_{narrow} \ell \bar{\nu}_{\ell})
= (0.18 \pm 0.02)\,\%
\label{eq:bsldpinarrow}
\end{equation}
is lower than the corresponding value for broad states,
obtained by averaging Belle \cite{Liventsev:2007rb}
and BaBar \cite{Aubert:2007qw, Aubert:2008ea} measurements:

\begin{equation}
BR(\bar{B}^0_d \to [D \pi]_{broad} \ell \bar{\nu}_{\ell})
= (0.42 \pm 0.06)\,\%
\label{eq:bsldpibroad}
\end{equation}

Computations of $BR(\bar{B} \to [D \pi]_{broad} \ell \bar{\nu}_{\ell})$
were done by several authors in the framework of heavy quark and chiral
symmetries \cite{Yan:1992gz, Lee:1992ih, Kramer:1992ag, Goity:1994xn}. 
They obtain a broad component which can be large but their predictions
vary over a wide range depending on their definition for the resonant
component and on the cut on the soft pion momentum. We have not used
their detailed expressions for the decay branching fraction and considered
that the contribution from the $D^*$ pole is dominant, as they had observed.
Our approach differs also because the coupling constant $g$ has now been accurately
measured and because we give a well defined scheme to compare experimental
measurements with theoretical predictions. 
We have found that the value expected for the $D^*_V$ component of the $D^*$ resonance is compatible 
with the $[D \pi]_{broad}$ measurements. Therefore, the broad contribution is perhaps neither,  as previously considered, the $D_0^*$ one, which should be very small, nor one coming from a radial excitation, as also suggested \cite{Bernlochner:2012bc}, but the $D^*_V$ one. An excess of events at low
$D \pi $ mass values is observed in Belle and BaBar analyses but the helicity
distribution measured by Belle does not favor the $D^*_V$ hypothesis. Meanwhile present statistics are too low to provide definite conclusions. Measurements of higher values for $\bar{B} \to D^* \ell^- \bar{\nu}_{\ell}$ branching fractions
obtained by fitting the inclusive lepton momentum distribution as compared with those obtained with exclusive analyses may point also to some missing $D^*_V$
component \cite{Aubert:2008yv,TheBABAR:2016lja}?

\subsection{$\bar{B} \rightarrow D^{*} \ell^- \bar{\nu}_{\ell}$ }
The semileptonic decay width for this reaction is used to measure the CKM 
parameter $\Vcb$ by comparing the corresponding experimental branching fraction 
with theoretical expectations, obtained in the hypothesis that the
$D^*$ is a stable particle.

Integrating over decay angles, the partial decay width depends on two variables:
$m_{D\pi}=\sqrt{s}$ and $w$, the latter being related to $q^2$, the invariant "squared mass"  of the two-lepton 
system:
\begin{equation}
w=\frac{m_B^2+s-q^2}{2 m_B \sqrt{s}}.
\end{equation}
If one assumes that the $D^*$ is stable, then $m_{D\pi}=m_{D^*}$ and the differential decay width becomes, in analogy to what was found in the nonleptonic case: 
\begin{equation}
\frac{d\Gamma}{dw}= \frac{G^2_F m^3_B}{48 \pi^3}r^3 \left (m_B-m_{D^*}\right )^2
\chi(w) \eta^2_{EW}{\cal F}^2(w) |V_{cb}|^2
\label{eq:dg1sl}
\end{equation}
where $r = m_{D^*}/m_B$.
The form factor ${\cal F}(w)$ depends on three form factors and is usually 
expressed in terms of one of them -$h_{A_1}(w)$- and of the ratios 
-$R_1(w),\,R_2(w)$- of the two others relative to $h_{A_1}(w)$.
\begin{eqnarray}
\chi(w) {\cal F}^2(w)&=& h^2_{A_1}(w)\sqrt{w^2-1}(w+1)^2 \left \{ 2 \left [ \frac{1-2wr+r^2}{(1-r)^2}\right ] \left [ 1 + R_1^2(w)\frac{w-1}{w+1}\right ] \right.\nonumber \\ 
& & \left. +  \left [ 1+(1-R_2(w))\frac{w-1}{1-r} \right ]^2 \right \}.
\label{eq:dgslw}
\end{eqnarray}
We use the parameterization of \cite{Caprini:1997mu} for the functions that enter
in Eq. (\ref{eq:dgslw}):
\begin{eqnarray}
h_{A_1}(w) &=&h_{A_1}(1)\left [ 1 -8\rho^2z+ (53 \rho^2-15)z^2-(231\rho^2 -91)z^3\right ],\nonumber\\
R_1(w) &=&R_1(1)-0.12 (w-1) + 0.05(w-1)^2, \nonumber\\
R_2(w) &=&R_2(1)+0.11 (w-1) - 0.06(w-1)^2
\label{eq:param_lellouch}
\end{eqnarray}
where $z=(\sqrt{w+1}-\sqrt{2})/(\sqrt{w+1}+\sqrt{2})$.
The values obtained by the HFAG group \cite{ref:hfag} for the parameters,
$\rho^2$, $R_1(1)$, and $R_2(1)$,
that enter in Eq. (\ref{eq:param_lellouch}) are the following:
\begin{equation}
\rho^2=1.205\pm0.026,\,R_1(1)=1.404\pm0.032,\,{\rm and}\,R_2(1)=0.854\pm0.020.
\end{equation}
They have been determined from a fit to experimental data that includes also 
the normalization for the decay rate:
\begin{equation}
\eta_{EW}{\cal F}(1) \Vcb=(35.61 \pm 0.43)\times 10^{-3}.
\end{equation}
Using these values, we have verified that, integrating Eq. (\ref{eq:dg1sl})
over $w$, to obtain the semileptonic decay partial width of the $\Bo$ meson,
we recover the central value of $4.88\,\%$ for the corresponding measured 
decay branching fraction which is given by HFAG.
Corresponding central values for the semileptonic decay width into a $D^*$, 
considered as a stable particle, are respectively for the neutral and the charged $B$ meson:
\begin{equation}
\Gamma_2^{sl}(B^0_d)=2.113\times 10^{-14}\,GeV\,{\rm and}\,\Gamma_2^{sl}(B^-)=2.116\times 10^{-14}\,GeV.
\label{eq:slwidth2}
\end{equation}
The small difference between these two values is attributed to differences
in the masses of the particles involved.

\subsubsection{Virtual D* contribution}
To evaluate the effects induced by the D* coupling to the
$D \pi$ final state, one muliplies  Eq. (\ref{eq:dg1sl})
 by:
\begin{equation}
R^i(s)= \frac{1}{\pi}\frac{\sqrt{s}\,\Gamma_i(s)}{(s-m^2_{D^*})^2+(\sqrt{s}\,\Gamma_{D^*}(s))^2}
\end{equation}

\noindent where the index "i" refers to the relevant decay channel.
In the limit $\Gamma_{D^*}(s)\to 0$, this expression corresponds to $\delta(s-m^2_{D^*})$ and one recovers Eq. (\ref{eq:dg1sl}) multiplied by the branching fraction
of the $D^*$ into the $i$ decay channel (${\cal B}_i=\Gamma_i(m_{D^*})/\Gamma_{D^*}(m_{D^*})$).
Total and partial widths include the Blatt-Weisskopf damping factor $F_R$. By
tradition, the $F_B$ damping term is not used when computing semileptonic
decays.

We have included three decay channels of the $D^*$: $D^0\pip$, $D^+\pi^0$, and
$D^+\gamma$ for the charged state and $D^0\pi^0$, $D^+\pi^-$, and
$D^0\gamma$ for the neutral one. If we integrate over $s$ and $w$, values 
for the 
semileptonic decay widths, $\Gamma_3^{sl}$, divided by $\Gamma_2^{sl}$,
are given in Table \ref{tab:widthbtodvlnu} (we have adopted the same notation
as for hadronic B decays: the index 2 refers to a stable $D^*$ particle):


\begin{table}[!htb]
\begin{center}
  \begin{tabular}{|c|c|c|c|c|c|}
    \hline
 $r_{BW}$\,$(GeV/c)^{-1}$  & 0  & 1  & 1.85 & 3 & 5 \\
\hline
$\Bob \to \Dstarp e^- \bar{\nu}_e$& $1.089$ & $1.072$ & $1.056$ & $1.041$ & $1.028$\\
$\Bm \to \Dstaro e^- \bar{\nu}_e$& $1.085$ & $1.068$ & $1.052$ & $1.038$ & $1.025$\\
\hline
  \end{tabular}
  \caption[]{\it {Partial decay widths for the channel $\bar{B}\to D^* e^- \bar{\nu}_e $ relative to the values obtained for a stable $D^*$ meson. The first 
line gives the value (in $(GeV/c)^{-1}$ units) of the parameter $r_{BW}$,
used in the Blatt-Weisskopf damping factor.}
  \label{tab:widthbtodvlnu}}
\end{center}
\end{table}

Depending on the value of the damping parameter, the semileptonic partial width
obtained by integrating over the $D^*$ mass distribution exceeds by 3 to 9 $\%$
the value obtained in the zero width approximation. This is a situation similar
to the decay $B \to \bar{D}^* \pi$ studied in Section \ref{sec:bneutral_dstar}.

The mass interval, centered on $m_{D^*}$, which is such that
the integral restricted over this interval is equal to $\Gamma_2^{sl}$
corresponds to $\Delta_3(m)=\pm (9-10)\,MeV/c^2$ and the obtained partial width is almost 
independent of the value of the damping parameter 
(with relative variations $<10^{-3}$). 

To obtain the value for $|V_{cb}|$ one needs the value of ${\cal F}(1)$
and this quantity
is evaluated for a stable $D^*$ particle. Therefore we consider that
theoretical expectations have to be compared with the measured branching
fraction restricted to the interval $\Delta_3(m)$. The event simulation
must not be used to correct for $D^*$ decays that are outside this
interval. 
One difficulty is
to fix the level of the combinatorial background under the $D^*$ signal,
in the $\Delta_3(m)$ mass interval,
because the $D^*$ signal is still present at large $D\pi$ masses and its exact
contribution depends on damping factors and on the opening of other
decay channels. It is therefore important to have a better experimental control
of the so called $D^{*}_V$ mass distribution. 
In present publications there is usually some missing information
to understand exactly how measurements were done. It would be nice if
the different experimental collaborations would clarify this situation. 

\subsection{$\bar{B} \rightarrow D^{*}_V \ell^- \bar{\nu}_{\ell}$ }

As we have noted, in previous sections, the $D^*_V$ component is not negligible in 
$\bar{B}_d^0 \rightarrow D^{0} \pi^+ \pi^-$ decays where it corresponds
to about $10\,\%$ of remaining events, once the $D^*$ peak is excluded. 
It is peaked at low
$D \pi$ mass values and it extends over a large mass range. 

To evaluate the branching fraction for 
$\bar{B}_d^0 \rightarrow D^{*+}_V \ell^- \bar{\nu}_{\ell}$
we have integrated the differential decay width
$d^2\Gamma/dw\,ds$ over $w$ and $s$ for $\sqrt{s}>m_{D^*}+9\,MeV/c^2$
(by comparison, in Belle, they select events with
$\sqrt{s}>m_{D^*}+1.5 \,MeV/c^2$ whereas  BaBar  uses   
$\sqrt{s}>m_{D}+180. \,MeV/c^2$).  


\begin{table}[!htb]
\begin{center}
  \begin{tabular}{|c|c|c|c|c|c|}
    \hline
  $r_{BW}$\,$(GeV/c)^{-1}$  & 0  & 1  & 1.85 & 3 & 5 \\
\hline
$\Bob \to D^{*+}_V e^- \bar{\nu}_e$& $0.48$ & $0.38$ & $0.29$ & $0.21$ & $0.14$\\
$\Bm \to D^{*0}_V e^- \bar{\nu}_e$&  $0.49$ & $0.39$ & $0.29$ & $0.21$ & $0.13$\\
\hline
  \end{tabular}
  \caption[]{\it {Estimated
 semileptonic branching fractions (in $\%$) for the channel
$\bar{B} \rightarrow D^{*}_V e^- \bar{\nu}_e$. The first line
gives the value (in $(GeV/c)^{-1}$ units) of the parameter $r_{BW}$,
used in the Blatt-Weisskopf damping factor.}
  \label{tab:btodvlnu}}
\end{center}
\end{table}

Therefore, comparing the values given in Table \ref{tab:btodvlnu} with the measurement
from Eq. (\ref{eq:bsldpibroad}), it appears that the $D^*_V$ component can 
explain all or a large fraction of the ``missing'' decay channel
in $\bar{B}_d^0 \to [D \pi]_{broad}] \ell \bar{\nu}_{\ell}$.

In addition, the $D^*_V$ component can be identified experimentally
because it has characteristic $m_{D\pi}$ (see Fig. \ref{fig:dpilnu}) and angular distributions.
Therefore, $B$ hadron semileptonic decays 
offer a nice
opportunity to study the $D\pi$ mass distribution of the $D^*_V$ component
because of the absence of any additional hadron in the decay final state. 
Such measurements can be considered at LHCb because present statistics from 
B-factories published analyses are too low for such studies. 
\begin{figure}[htbp]
  \begin{center}
     \epsfig{file=
     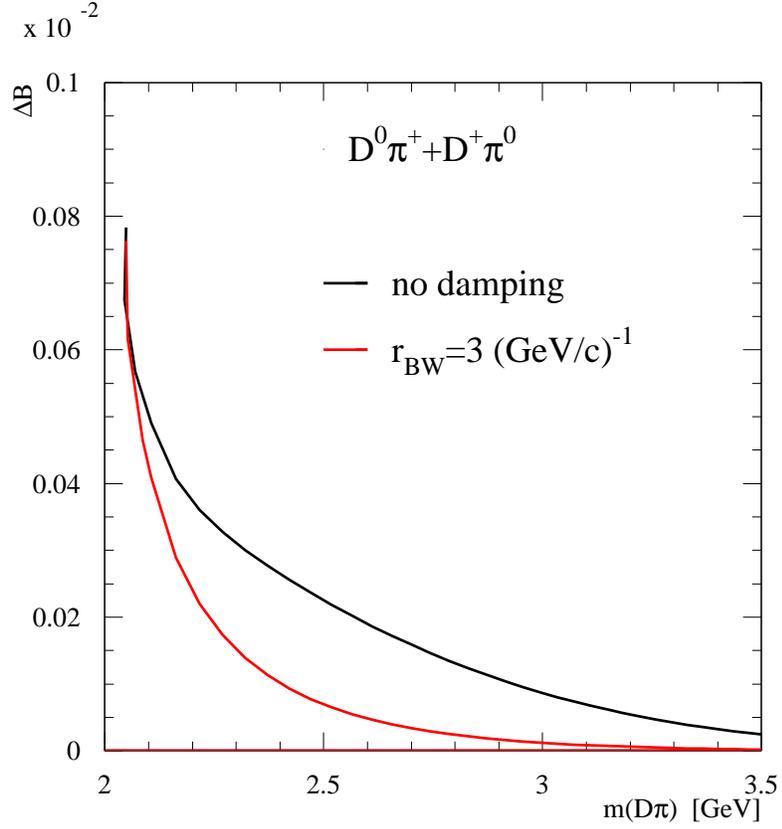,width=\textwidth} 
  \end{center}
 \caption[]{\it Expected $D\pi$ mass distribution for the $D^*_V$ component in $\bar{\Bo}$ hadron semileptonic decays. The two curves, correspond to expectations without damping (black) and with (red) using the Blatt-Weisskopf distribution
with $r_{BW}=3 \, (GeV/c)^{-1}$. $\Delta B$ is the branching fraction expected in  each $m_{D\pi}^2$ bin. There are 100 equal-size bins between $(m_{D^{*+}}+\Delta_3)^2$ and $m_{B^0_d}^2$.
}   
\label{fig:dpilnu}
\end{figure}


\section{Conclusions}
We have found that, to compare expected branching fractions with experiment
in $\bar{B} \to D^* \pi$ and $\bar{B} \to D^* \ell \bar{\nu}_{\ell}$ 
decays, that are always provided from theory in the zero width limit, one
has to integrate the $D^* \to D \pi$ mass distribution from threshold
up to $m^0_{cut} = m_{D^*}+ (9-10)\,MeV/c^2$ (Eq. \ref{eq:mcut}); 
this interval corresponds
to more than one hundred times the intrinsic resonance width. 
In this way, the two values
are expected to agree at the permil level, independently of effects
from damping factors that are usually introduced in decay 
amplitudes (see Table \ref{tab:dstar}). Such
an accuracy supposes a precise control of the $D \pi$ combinatorial
background level, within the selected mass range. This is a priori non trivial 
because of the presence of $D^*$ decays at high mass values (the so-called 
$D^*_V$ events) that need to be estimated. Therefore it is also important to have a good
understanding of this component.

The $D^*_V$ component corresponds to $m(D\pi)>m^0_{cut}$ and comes mainly
from the real part of the $D^*$ propagator. We have shown that its
relative importance, when compared with the zero width limit, is essentially
independent of the value of the vector resonance intrinsic width
(see section \ref{sec:fixed_coupling}) when this
quantity is computed according to $\Gamma_0 \propto g^2 p_0^{*3}$, the 
coupling $g$ being a constant fixed by strong interactions. This result
is verified by changing the mass of an hypothetical vector resonance, decaying 
into $D\pi$, between threshold and $2.1\,GeV/c^2$, which corresponds to
$\Gamma_0$ varying between $\sim 0$ and $7\,MeV$. We find that
the measured $D^*_V$ production rate is compatible with expectations
obtained from the $D^*$ within uncertainties that are quite large at present
(see Table \ref{tab:b0todpipi}).
Specifically, predicted branching fractions depend on the parameterization
of damping factors and on the way they are computed (see Table \ref{tab:dstarv}). 
In this note we have not
really addressed some aspects that can still affect the $D^*_V$ evaluation, such
as: the physical origin and interpretation of damping factors, and the opening
of new decay channels at large masses.

We therefore consider that it is important to have an experimental control
of the $D^*_V$ component. The $B^- \to D^+ \pi^- \pi^-$ seems promising
in this respect because large statistics can be analyzed and
no resonance is expected in the two-pion channel (see Section \ref{sec:bcharged}). 
Another appealing
possibility is the semileptonic $\bar{B} \to D \pi \ell \bar{\nu}_{\ell}$
decay because of the absence of a third hadron in the final state and 
because the $D^*_V$ component is expected to dominate the $D\pi$ channel
(see Section \ref{sec:bsemilept}).

\section*{Acknowledgements}
We would like to thank V. Tisserand and Wenbin Qian for providing us with some details about
the $\Bo \to \Dob \pim \pip$ decay measurement in LHCb. We thank in particular T. Gershon and T. Latham for answering several questions we had on $\bar{B} \to D \pi\pi$ analyses in BaBar. Comments from B. Kowalewski on a possible $D^*_V$ contribution to explain the difference between inclusive and exclusive measurements
of the $\bar{B} \to D^* \ell^- \bar{\nu}_{\ell}$ branching fraction are included. We have also commented
searches for the $D^*_V$ component in exclusive semileptonic decays done in
BaBar and Belle experiments, as suggested by M. Rotondo.

\section*{Appendix 1: An illustrative model} 


We use a simplified model to display the physical origin
of the relatively large difference existing between the partial decay widths $\Gamma_3$
and $\Gamma_2 \times BR$, in spite of the extreme smallness of $\Gamma_{{D^*}}$.
The main simplifications are that\\
\begin{itemize} \item[1)] we discard any damping factor; \item[2)] in the denominator of the  propagator, we consider a fixed width $\Gamma_{D^*}$ depending only on the $D^*$ mass instead of  $\Gamma(s)$, which would be more respectful of unitarity.\end{itemize} Differences between this simplified model and
numerical results quoted in the article, that were obtained using a variable
decay width, do not seem essential. The magnitude of the effect is the same as in the more complete calculation.

The expression for $\Gamma_3$ is given in Eq. (\ref{gammatot}). It can be rewritten:
\begin{equation}
\Gamma_3=C \int^{(m_B-m_{\pi})^2}_{(m_D+m_{\pi})^2} ds\,\,\frac{\phi(s)}{(s-m^2_{D^*})^2+ (m_{D^*}\Gamma_{D^*})^2}
\label{gammatot22}
\end{equation}

The constant $C$ is some combination of numerical factors and coupling constants not relevant here since we discuss only ratios. The function $\phi(s)$ is equal to:
\begin{equation}
\phi(s)=\frac{p^{\prime\,3}_2(s) \,p_1^3(s)}{s^{3/2}}.
\label{gammaphi}
\end{equation}

In the zero $D^*$ width limit one gets the equality\footnote{We recall that $BR$ stands for $BR_{D^{*-}\to \bar{D}^0 \pi^-}({m^2_{D^*}}).$}:
\begin{equation}
\lim_{\Gamma_{D^*}\to 0} \Gamma_3= \Gamma_2 \times BR = \nonumber 
C \pi \phi(m^2_{D^*})/(m_{D^*}\Gamma_{D^*})
\end{equation}

The intermediate expression in the equation above corresponds to  $\Gamma_{\Bo\to D^{*-} \pi^+}\times \,BR$ in Eq. (\ref{eq:limgamma3}).
 
To display the difference between $\Gamma_3$ and its limit, one may rewrite  $\pi\phi(m^2_{D^*})/(m_{D^*}\Gamma_{D^*})$ as the integral:
\begin{equation}
 \int^{(m_B-m_{\pi})^2}_{(m_D+m_{\pi})^2} ds\,\,\frac{\phi(m^2_{D^*})}{(s-m^2_{D^*})^2+ (m_{D^*}\Gamma_{D^*})^2}
\end{equation}
This is an approximation, but a very good one. It amounts to replace $\frac{\pi}{m_{D^*}\Gamma_{D^*}}$ (which is the exact result for the same integral, but with infinite bounds) by:
\begin{eqnarray}
\frac{\pi}{m_{D^*}\Gamma_{D^*}}\left [ 1 - \frac{m_{D^*}\Gamma_{D^*}}{\pi\,A } - \frac{m_{D^*}\Gamma_{D^*}}{\pi\, B}\right ]
\end{eqnarray}
where $A=m^2_{D^*}-(m_D+m_{\pi})^2$
and $B=(m_B-m_{\pi})^2-m^2_{D^*}$.


The relative difference between the $\Gamma_3$ and 
$\Gamma_2 \times BR$
decay widths is then equal to: 
\begin{equation}\begin{split}
R-1&=\frac{\Gamma_3-\Gamma_2 \times BR}{\Gamma_2 \times BR}\\&\simeq  m_{D^*}\Gamma_{D^*}\frac{1}{\pi \phi(m^2_{D^*})}\int^{(m_B-m_{\pi})^2}_{(m_D+m_{\pi})^2} ds\,\,\frac{\phi(s)-\phi(m^2_{D^*})}{(s-m^2_{D^*})^2+ (m_{D^*}\Gamma_{D^*})^2}
\end{split}\end{equation}

In the factors in front of the integral, the critical dependence of $\Gamma_{D^*}$ on the $m_{D^*}$ mass through the factor $p_1^3(m^2_{D^*})$ (see Eq. (\ref{eq:width_no_p})) which could lead one to believe that the expression tends to zero at threshold, is compensated exactly by the same factor in $\phi(m^2_{D^*})$. Displaying explicitly the coupling constant factor, and taking into account that the above expression involves the total width $\Gamma_{D^*}$
instead of the partial one given in Eq. (\ref{eq:width_no_p}) ($BR \approx 2/3$), one ends with:
\begin{eqnarray}
R-1 \approx \frac{1}{16 \pi} g^2_{D^*D\pi} m_{D^*}^2 \frac{1}{\pi}\frac{1}{p^3_2 (m^2_{D^*})} \int ds \frac {\phi(s)-\phi(m^2_{D^*})} 
{(s-m^2_{D^*})^2+(m_{D^*} \Gamma_{D^*})^2}, 
\end{eqnarray}

One sees that there remains only a smooth dependence on $m_{D^*}$ in $p_2^3(m^2_{D^*})$ ($B$ decay) and from the integral, as observed in the numerical curves of subsection 3.2.
In particular, the limit $m_{D^*} \to m_D+m_{\pi}$ is finite and $\neq 0$, although the width goes to $0$. 

The magnitude of $R-1$ is controlled by the magnitude of the coupling constant: 
 it is roughly proportional to $g^2_{D^*D\pi}$}. One would recover $R \approx 1$ if this coupling was very small, as shown in the beginning of the paper. But of course it is not small in reality. The smallness of the $D^*$ width is accidental, only due to the proximity to the threshold, and the coupling is comparable to the one for other strong couplings like $g_{NN\pi}$. Numerically,
one finds  for the physical value of $g^2_{D^*D\pi} $ and for $m_{D^*}$ very close to the threshold, in fact for an arbitrarily small width, $R-1 \approx 0.09$.



\section*{Appendix 2: The Determination of 
the $g_{D^*D\pi}$ coupling constant}

The value of $g_{D^{*-}\bar{D}^0\pim}$ is obtained from the measurement of the 
$\Dstarp$ hadronic decay width using the expression given in
Eq. (\ref{eq:width_no_p}).

Experiments have measured the total width of the $\Dstarp$ meson and 
the small contribution from electromagnetic decays needs to be subtracted
to obtain the hadronic component.
Using values quoted in \cite{Patrignani:2016xqp} this gives:
\begin{equation}
\left.\Gamma \right|_{\Dstarp \to D\pi}^{expt.} = (83.4 \pm 1.8) \times (1- 0.016\pm0.004)keV = (82.1 \pm 1.8)keV. 
\end{equation}
It corresponds to:
\begin{equation}
g_{D^*D\pi}^{expt.}= 16.81 \pm 0.18.
\end{equation}
This value is obtained using the hypothesis of I-spin symmetry to relate
the $\Do \pip$ and $\Dp \pio$ decay channels of the $\Dstarp$, taking into account the difference of the decay momenta.
The validity of this hypothesis can be checked by comparing the 
measured $(67.7\pm 0.5)\%$ and expected $(67.6\pm 0.3)\%$ values 
for $BR_{\Dstarp \to \Do \pip}$.

\section*{Appendix 3: On the {\it s}-dependence of the imaginary part of  the resonance propagator}

In this Appendix we want to demonstrate the statements and claims formulated 
in the text about the $s$ dependence of  $-i \sqrt{s}~\Gamma(s)$, i.e. of the imaginary part of the self-energy, namely that it is proportional to $\frac{1}{\sqrt{s}}$ times the usual factors $q(s)^{2 l+1}$, for an $l$ decay partial wave\footnote{This differs from the expressions given in the new section "48. Resonances"  in the 2018 edition of PDG \cite{Patrignani:2016xqp}, Eqs. (48.22) and (48.23), where it is only $\propto q(s)^{2 l+1}$.}, $q(s)$ being the decay momentum for a particle of mass $\sqrt{s}$ in its rest frame. This conclusion is obtained by using a loop model.


\subsection*{\small Basis of the calculation}
 For  the sake of simplicity, we shall make our demonstration for the case of a scalar resonance of mass $M$ decaying into two identical scalars of mass $m$.
The extension to the case of a vector resonance like the $D^*$ is straightforward.

The propagator can be written as:
\begin{eqnarray}
\frac{1} {s-M^2-\Sigma(s)}
\end{eqnarray}
with $s=p^2$ and $\Sigma(s)$ is the self energy contribution, having in fact obviously a dimension $mass~squared$. We are here only interested in the imaginary part of $\Sigma(s)$, although there is of course also a real $s$-dependent mass shift. In the literature, this imaginary (absorptive) part is either denoted as $-i \sqrt{s}~\Gamma(s)$  or $-i M \Gamma_M(s)$. Note that in the beginning this is only a matter of convention, if the quantities $\Gamma_M(s)$ or $ \Gamma(s)$ are evaluated accordingly, with M $\Gamma_M(s)=\sqrt{s}~\Gamma(s)$, from the same imaginary part, but it may be a source of confusion. The justification of such notations is just to explicit the dimension of a $mass~squared$ and to recall the relation with the physical width, let us say, which would be defined at the pole mass. In fact, as is obvious in the calculation, the absorptive cannot depend on the mass $M$, but only on $s$ and $m$, so that the notation  $-i M \Gamma(s)$ introduces a dependence on $M$ in both factors which is rather artificial. We therefore stick to  the notation $-i \sqrt{s}\,\Gamma(s)$.

Now the only simple way to get an explicitly covariant expression for $-i \sqrt{s}\,\Gamma(s)$ in accordance with the analyticity and unitarity requirements of field theory is to use Feynman diagrams and to generate the self-energy and its  imaginary part by the loop contributions, which are produced by iteration of the (normal ordered) coupling of the $M$ particle to the two identical others:
 \begin{eqnarray}
\lambda~:\Phi_M~(\phi_m)^2:
\end{eqnarray}
with $\lambda$ the (dimensionful) coupling constant to appear also in the decay width, and $\Phi_M,\phi_m$  the fields corresponding to the respective scalar particles.
We therefore proceed by applying the standard Feynman rules to the calculation of the corresponding propagator. 


\subsection*{\small Calculation}
Let $P$ be the momentum entering the loop and $s\,=\,P^2.$ The self-energy is generated by the series of loop diagrams: 
\begin{eqnarray}
\frac {i}{s-M^2} \sigma ~\frac {i}{s-M^2} + \frac {i}{s-M^2} \sigma \frac {i}{s-M^2}\sigma ~\frac {i}{s-M^2} +...
\end{eqnarray}
with $\sigma$ coming from the loop integral: 
\begin{eqnarray}
\sigma= \frac{i^2 \lambda^2}{2} \int \frac {d^4 K}{(2 \pi)^4} \frac {i^2}{(K^2-m^2)((P-K)^2-m^2)}
\end{eqnarray}
The factors in front of the integral come from twice the vertex $i \lambda$, and a factor $1/2$ for the bosonic loop.
These loop contributions, added to the bare $i \frac {1}{s-M^2}$, give:
\begin{eqnarray}
\frac {i}{s-M^2-\Sigma(s)},
\end{eqnarray}
with $\Sigma=i \sigma(s)$, whence finally:
\begin{eqnarray}
\Sigma= i\frac{\lambda^2}{2} \int \frac {d^4 K}{(2 \pi)^4} \frac {1}{(K^2-m^2)((K-P)^2-m^2)}
\end{eqnarray}
It is obvious 
that $\Sigma$ does not contain any dependence on the mass $M$ of the decaying scalar. 

We need only the absorptive part of $\Sigma$, which we obtain by means of the Cutkosky rule, i.e. the substitution:
\begin{eqnarray}
\frac {1}{u-m^2} \to 2 \pi i \delta(u-m^2)
\end{eqnarray}
for each denominator $u-m^2$ inside the loop.
The calculation is easily done in the frame where $\vec{p}=0$, yielding:
\begin{eqnarray}
-2 i\sqrt{s}~\Gamma(s)=Disc~\Sigma=\frac{i \lambda^2} {2} \left(-\frac {1}{4\pi} \frac{q(s)} {\sqrt{s}}\right) \\
\Gamma(s)=\frac{1}{2}\frac{\lambda^2}{8 \pi} \frac{q(s)}{s}
\end{eqnarray}
with $q(s)=\frac {1}{2} \sqrt{s-4m^2}$, i.e. the decay momentum of the particle of mass $\sqrt{s}$  into two decay products with equal mass $m$ in its rest frame.



\begin{thebibliography}{20}



\bibitem{Neubert:1997uc}
Matthias Neubert and Berthold Stech.
\newblock {Nonleptonic weak decays of B mesons}.
\newblock {\em Adv. Ser. Direct. High Energy Phys.}, 15:294--344, 1998.

\bibitem{Lichard:1998ht}
Peter Lichard.
\newblock {Are the production and decay of a resonance always independent?}
\newblock {\em Acta Phys. Slov.}, 49:215--230, 1999.

\bibitem{Isgur:1988vm}
Nathan Isgur, Colin Morningstar, and Cathy Reader.
\newblock {The a1 in tau Decay}.
\newblock {\em Phys. Rev.}, D39:1357, 1989.


\bibitem{weinstein:1999}
Alan Weinstein.
\newblock {Breit Wigners and Form Factors}
\newblock {CBX 99-55 (Unpublished report) September 1999}

\bibitem{Zemach:1965zz}
Charles Zemach.
\newblock {Determination of the Spins and Parities of Resonances}.
\newblock {\em Phys. Rev.}, 140:B109--B124, 1965.

\bibitem{Zemach:1968zz}
Charles Zemach.
\newblock {Use of angular momentum tensors}.
\newblock {\em Phys. Rev.}, 140:B97--B108, 1965.

\bibitem{Patrignani:2016xqp}
C.~Tanabashi et~al.
\newblock {Review of Particle Physics}.
\newblock {\em Phys.\ Rev.\ D}, {\bf 98}, 030001 (2018).

\bibitem{ref:btod0-belle}
  A.~Kuzmin {\it et al.} [Belle Collaboration],
  Phys.\ Rev.\ D {\bf 76} (2007) 012006.


\bibitem{ref:btodstar_babar_1}
B.~Aubert {\it et al.} [BaBar Collaboration],
  Phys.\ Rev.\ D {\bf 75}, 031101 (2007).


\bibitem{ref:btodstar_babar_2}
B.~Aubert {\it et al.} [BaBar Collaboration],
  Phys.\ Rev.\ D {\bf 74}, 111102 (2006).


\bibitem{ref:bmtodp-belle}
 K.~Abe {\it et al.} [Belle Collaboration],
  Phys.\ Rev.\ D {\bf 69} (2004) 112002
  doi:10.1103/PhysRevD.69.112002
  [hep-ex/0307021].


\bibitem{ref:btod0-babar}
P.~del Amo Sanchez {\it et al.} [BaBar Collaboration],
  PoS ICHEP {\bf 2010} (2010) 250
  [arXiv:1007.4464 [hep-ex]].


\bibitem{ref:btod0-lhcb}
R.~Aaij {\it et al.} [LHCb Collaboration],
  Phys.\ Rev.\ D {\bf 92} (2015) 3,  032002
  doi:10.1103/PhysRevD.92.032002
  [arXiv:1505.01710 [hep-ex]].


\bibitem{ref:bmtodp-lhcb}
  R.~Aaij {\it et al.} [LHCb Collaboration],
  Phys.\ Rev.\ D {\bf 94} (2016) no.7,  072001
  doi:10.1103/PhysRevD.94.072001
  [arXiv:1608.01289 [hep-ex]].


\bibitem{ref:bmtodp-babar}
B.~Aubert {\it et al.} [BaBar Collaboration],
  Phys.\ Rev.\ D {\bf 79} (2009) 112004
  doi:10.1103/PhysRevD.79.112004
  [arXiv:0901.1291 [hep-ex]].



\bibitem{ref:bdpidl_th}

V. Morenas, A. Le Yaouanc, L. Oliver, O. Pene, J.C. Raynal, Phys.Rev. D56 (1997) 5668-5680, hep-ph/9706265 

See also the more recent discussion and references in 
Alain Le Yaouanc, Olivier Pène, Int.J.Mod.Phys. A30 (2015) no.10, 1543009,
arXiv:1408.5104 [hep-ph]

\bibitem{Liventsev:2007rb}
  D.~Liventsev {\it et al.} [Belle Collaboration],
  Phys.\ Rev.\ D {\bf 77} (2008) 091503.

\bibitem{Aubert:2007qw}
  B.~Aubert {\it et al.} [BaBar Collaboration],
  Phys.\ Rev.\ Lett.\  {\bf 100} (2008) 151802.

\bibitem{Aubert:2008ea}
  B.~Aubert {\it et al.} [BaBar Collaboration],
  Phys.\ Rev.\ Lett.\  {\bf 101} (2008) 261802.

\bibitem{Isgur:1991}
  N. Isgur and M.Wise, Phys. Rev. D  {\bf 43} (1991) 819.




\bibitem{Yan:1992gz} 
  T.~M.~Yan, H.~Y.~Cheng, C.~Y.~Cheung, G.~L.~Lin, Y.~C.~Lin and H.~L.~Yu,
  Phys.\ Rev.\ D {\bf 46}, 1148 (1992)
  Erratum: [Phys.\ Rev.\ D {\bf 55}, 5851 (1997)].
  doi:10.1103/PhysRevD.46.1148, 10.1103/PhysRevD.55.5851

\bibitem{Gourdin} 
  Gourdin,~M. and Salin,~Ph.
  Nuovo Cimento 27 (1963) 193
    

\bibitem{Lee:1992ih} 
  C.~L.~Y.~Lee, M.~Lu and M.~B.~Wise,
  Phys.\ Rev.\ D {\bf 46}, 5040 (1992).
  doi:10.1103/PhysRevD.46.5040

\bibitem{Kramer:1992ag} 
  G.~Kramer and W.~F.~Palmer,
  Phys.\ Lett.\ B {\bf 298}, 437 (1993).
  doi:10.1016/0370-2693(93)91847-G


\bibitem{Caprini:1997mu} 
  I.~Caprini, L.~Lellouch and M.~Neubert,
  Nucl.\ Phys.\ B {\bf 530}, 153 (1998)
  doi:10.1016/S0550-3213(98)00350-2
  [hep-ph/9712417].
  
\bibitem{Goity:1994xn}
 J.~L.~Goity and W.~Roberts, 
Soft pion emission in semileptonic B meson decays,
  Phys.\ Rev.\ D {\bf 51} (1995) 3459,
  doi:10.1103/PhysRevD.51.3459
  [hep-ph/9406236].


\bibitem{Bernlochner:2012bc} F.~U.~Bernlochner, Z.~Ligeti and S.~Turczyk, 
A Proposal to solve some puzzles in semileptonic B decays,
  Phys.\ Rev.\ D {\bf 85} (2012) 094033,
  doi:10.1103/PhysRevD.85.094033,
  [arXiv:1202.1834 [hep-ph]].

\bibitem{Aubert:2008yv}
  B.~Aubert {\it et al.} [BaBar Collaboration],
  Phys.\ Rev.\ D {\bf 79} (2009) 012002

\bibitem{TheBABAR:2016lja}
  J.~P.~Lees {\it et al.} [BaBar Collaboration],
  Phys.\ Rev.\ D {\bf 95} (2017) no.7,  072001.


\bibitem{ref:hfag}
Y.~Amhis {\it et al.},
  arXiv:1612.07233 [hep-ex].



\end{thebibliography}

\end{document}